\documentclass[sigconf,authorversion,nonacm]{cidr-2024}   % use this for arxiv submission

\usepackage{subfiles}
\usepackage{enumitem}
\usepackage{setspace}
\usepackage[ruled, vlined, linesnumbered]{algorithm2e}
\PassOptionsToPackage{noend}{algpseudocode}
\usepackage{algpseudocode}
\usepackage{subfig}
\usepackage{listings}
\usepackage{balance}
\usepackage{xr}
\usepackage{caption,tcolorbox,xcolor}
\usepackage[normalem]{ulem}
\usepackage{tabularray}
\UseTblrLibrary{booktabs}
\usepackage{multirow}

\begin{document}

\title{Enhancing Computation Pushdown for Cloud OLAP Databases}

\author{Yifei Yang}
\affiliation{%
  \institution{University of Wisconsin-Madison}
}
\email{yyang673@wisc.edu}

\author{Xiangyao Yu}
\affiliation{%
  \institution{University of Wisconsin-Madison}
}
\email{yxy@cs.wisc.edu}

\author{Marco Serafini}
\affiliation{%
  \institution{University of Massachusetts-Amherst}
}
\email{marco@cs.umass.edu}

\author{Ashraf Aboulnaga}
\affiliation{%
  \institution{University of Texas at Arlington}
}
\email{ashraf.aboulnaga@uta.edu}

\author{Michael Stonebraker}
\affiliation{%
  \institution{Massachusetts Institute of Technology}
}
\email{stonebraker@csail.mit.edu}

%% customized commands
\newcommand{\yxy}[1]{\textcolor{blue}{yxy: #1}}
\newcommand{\yifei}[1]{\textcolor{purple}{yifei: #1}}
\newcommand{\ms}[1]{\textcolor{red}{#1}}
\newcommand{\msc}[1]{\textcolor{red}{[marco: #1]}}
\newcommand{\aia}[1]{\textcolor{brown}{ashraf: #1}}

\newcommand{\name}{FPDB\xspace}
\newcommand{\np}{\textit{No pushdown}\xspace}
\newcommand{\ep}{\textit{Eager pushdown}\xspace}
\newcommand{\ap}{\textit{Adaptive pushdown}\xspace}
\newcommand{\spara}[1]{\vspace*{0.05in} \noindent \textbf{#1.}}
\newcommand{\sspara}[1]{\vspace*{0.05in} \noindent \textit{#1.}}
\newcommand{\texttwin}[2]{\textit{#1}$_{\textit{#2}}$}
\newcommand{\eqtwin}[2]{#1_{\textit{#2}}}
\newcommand{\algsp}{\hspace{0.1em}}

\captionsetup[figure]{font=small}
\captionsetup[table]{font=small}
\captionsetup[lstlisting]{font=small}
\SetAlCapFnt{\small}

\setlength{\dbltextfloatsep}{0.3cm}
\setlength{\textfloatsep}{0.3cm}
\setlength{\floatsep}{0.2cm}
\setlength{\intextsep}{0.2cm}

\begin{abstract}
Network is a major bottleneck in modern cloud databases that adopt a storage-disaggregation architecture. Computation pushdown is a promising solution to tackle this issue, which offloads some computation tasks to the storage layer to reduce network traffic. Existing cloud OLAP systems statically decide whether to push down computation during the query optimization phase and do not consider the storage layer’s computational capacity and load. Besides, there is a lack of a general principle that determines which operators are amenable for pushdown. Existing systems design and implement pushdown features empirically, which ends up picking a limited set of pushdown operators respectively.

In this paper, we first design \ap as a new mechanism to avoid throttling the storage-layer computation during pushdown, which pushes the request back to the computation layer at runtime if the storage-layer computational resource is insufficient. Moreover, we derive a general principle to identify pushdown-amenable computational tasks, by summarizing common patterns of pushdown capabilities in existing systems. We propose two new pushdown operators, namely, \textit{selection bitmap} and \textit{distributed data shuffle}. Evaluation results on TPC-H show that \ap can achieve up to 1.9$\times$ speedup over both \np and \ep baselines, and the new pushdown operators can further accelerate query execution by up to 3.0$\times$.
\end{abstract}

\maketitle

\section{Introduction} \label{sec:intro}

Modern OLAP databases are aggressively moving to the cloud for high elasticity and low cost. These systems adopt a storage-disaggregation architecture which  manages computation and storage into separate layers. The network between the two layers has become a major performance bottleneck %for analytical queries 
due to its relatively lower bandwidth and higher latency \cite{clouddb}. \textit{Computation pushdown} is a promising solution to mitigate the network bottleneck, where some computation logic is sent and evaluated close to the storage, resulting in less data returned to the computation layer~\cite{pushdowndb, fpdb, s3select, spectrum, presto}. %PushdownDB \cite{pushdowndb} and FlexPushdownDB (FPDB) \cite{fpdb} leverage S3 Select \cite{s3select} to push selection, projection, and simple aggregation to the S3~\cite{s3} object storage. Redshift Spectrum \cite{spectrum} further supports group-by clauses in the pushdown query.

In existing cloud OLAP DBMSs, the query execution engine decides whether to push down computation %occurs with the execution of a query is decided by the databased engine 
during the query optimization phase. % --- a portion of the query plan generated from the optimizer is offloaded to the storage layer. 
For example, Presto~\cite{presto} enables pushdown for all filter operators to S3 by setting a flag in the configuration file (\textit{"hive.s3select-pushdown.enabled=true"}). 
However, pushdown in existing systems does not consider %are offloaded to storage without considering 
the current storage layer’s computational capacity and load at the time when the query is executed. If the storage-layer computation resource is scarce (e.g., due to multi-tenancy), computation pushdown may hurt performance for a particular query. Unfortunately, it is difficult and sometimes impossible to predict the storage-side computational load ahead of query execution. 
%The runtime resource utilization in the storage layer may render pushdown futile, which is hard to predict ahead of execution. 
Another limitation of existing pushdown systems is the \emph{lack of a general principle that determines which operators are amenable for pushdown}. 
Existing systems empirically consider only simple operators such as 
% guides the decision Moreover, the implementation of pushdown functionalities in existing cloud OLAP DBMSs is mostly empirically based. For example, simple pushdown operators, such as 
selection, projection, and aggregation due to their simplicity and effective traffic reduction. 
We believe a larger set of operators can benefit from pushdown and a principle should exist to decide which operators should be considered. 
%, are widely endorsed due to the ease of implementation and promising potential network traffic reduction. Intuitively, operators that are amenable to pushdown may share a common pattern, which can be used to discover additional operators that can benefit from pushdown.

In this paper, we address both limitations above. % in existing pushdown systems. 
%\yxy{Should discuss adaptive pushdown first in the following paragraph.}
To address the first limitation, we propose an adaptive query processing approach that adapts the query plan while the query is executed to consider the current load on storage nodes.
We first explore the design space and analyze the theoretical bound --- what is the optimal division of the computation tasks between pushdown and non-pushdown to achieve the best overall performance. Then we design \ap as a new mechanism to avoid throttling the storage-layer computation during pushdown. Instead of having the database engine making the pushdown decisions, \ap lets the storage layer to decide whether to execute an incoming pushdown request, or to push the request back to the computation layer. When a pushback happens, the computation layer reads the raw data from the storage layer and processes the task locally. Intuitively, pushdown requests should be executed at the storage layer when sufficient computation resources exist, and pushed back when the storage layer is saturated. We will demonstrate that the proposed mechanism can perform close to the theoretical bound.
% We construct a lightweight model to intelligently make this decision at runtime in the storage layer. 
Evaluation results show that \ap outperforms both \np and \ep baselines by up to 1.9$\times$ on the TPC-H~\cite{tpch} benchmark.

% Besides proposing a framework that works in practice, we further explore the design space and analyze the theoretical bound --- what is the optimal division of the computation tasks between pushdown and non-pushdown to achieve the best overall performance. We will demonstrate that the proposed practical algorithm can perform close to the theoretical bound. %The theoretical optimal algorithm itself cannot be directly deployed, but provides vital insights about the effectiveness of the mechanism of adaptive pushdown.

To address the second limitation, we derive a general principle to identify pushdown-amenable computational tasks, % what computation should be considered for pushdown to the storage layer, 
through summarizing common patterns of pushdown capabilities in existing systems. First, pushdown tasks should be \textit{local} --- the pushdown computation should access data only within a single storage node
% \msc{we have not defined what a partition is yet}
and not incur data transfer within the storage layer. Second, pushdown tasks should be \textit{bounded} --- 
%Namely, the pushdown computation should not incur data transfer within the storage layer (i.e., \textit{local}) and 
a pushdown task should require at most linear CPU and memory resources with respect to the accessed data size. This principle preserves the benefits of storage-disaggregation, and simplifies resource isolation and security in a multi-tenant environment. Following the principle above, we %explore the design space and 
further identify two operators that can benefit from pushdown to the storage layer --- \textit{selection bitmap} and \textit{distributed data shuffle}.
Evaluation results show that the two new pushdown operators can further accelerate end-to-end query processing on TPC-H by up to 3.0$\times$ and 1.7$\times$ respectively.

\begin{itemize}

\item We develop \ap to leverage storage-layer computation dynamically, which utilizes a pushback mechanism to determine whether a pushdown operator should be executed in the storage.

\item We infer a general principle to determine whether a database operator is amenable to pushdown from existing cloud OLAP DBMSs, and identify two more common operators that can benefit from being offloaded to the storage layer --- \textit{distributed data shuffle} and \textit{selection bitmap}.

\item We implement and integrate \ap and new pushdown operators into FlexPushdownDB (\name)~\cite{fpdb}, an open-source C++-based cloud-native OLAP DBMS.
% \aia{You need to talk about FBDB when you first introduce it} We show the performance of \ap and new pushdown operators and further compare  \ap with the theoretical optimal bound.
% Evaluation shows that pushdown of these new operators leads to 46\% speedup on TPC-H benchmark, and 2.1$\times$ on the Star Schema Benchmark. Adaptive pushdown can achieve up to 3.9$\times$ speedup compared to eager pushdown when the storage layer computational resource is scarce.

\end{itemize}

The rest of the paper is organized as follows. Section~\ref{sec:back} introduces the background of pushdown in existing cloud OLAP DBMSs and illustrates the motivations. Section~\ref{sec:adapt} demonstrates adaptive pushdown and analyzes its theoretical optimum. In Section~\ref{sec:op}, we present a general principle of pushdown that is summarized from the implementation of existing cloud OLAP DBMSs, and identify new pushdown operators. Section~\ref{sec:impl} reveals additional implementation details.  Section~\ref{sec:eval} evaluates the performance of both adaptive pushdown and new pushdown operators using \name. Finally, Section~\ref{sec:related} discusses the related work and Section~\ref{sec:conclu} concludes the paper. 

\section{Background and Motivation} \label{sec:back}

This section describes the background of computation pushdown (Section~\ref{sec:back-pushdown}) and limitations in existing cloud OLAP databases that support computation pushdown (Section~\ref{sec:back-limit}).

\subsection{Computation Pushdown} \label{sec:back-pushdown}

The concept of computation pushdown was incubated in database machines since the 1970s. The early systems push computation to storage via special hardware. Database machines like the Intelligent Database Machine~\cite{idm}, Grace~\cite{grace}, IBM Netezza data warehouse appliances~\cite{netezza}, and Oracle Exadata Database Machine~\cite{exadata} move simple functionalities including filtering and projection close to disks. Other research areas, such as Smart Disks/SSD~\cite{do2013query, hrl, biscuit, summarizer, ibex, aquoman} and processing-in-memory~\cite{ghose2018enabling, kepe2019database} also endorse this spirit.

Cloud databases have emerged in the last decade, with computation and storage disaggregated, especially for analytical queries. The disaggregated architecture supports certain amount of \textit{computation within the storage layer}, so that some operators can be offloaded to storage to reduce network traffic. For example, S3 Select~\cite{s3select} is a feature inside the S3~\cite{s3} storage layer to execute pushdown tasks including selection, projection, and aggregation. The actual computation can happen either on the storage servers (e.g., Aurora~\cite{aurora1, aurora2}), or in a different sub-layer close to the storage devices (e.g., S3 Select~\cite{s3select}, Redshift Spectrum~\cite{spectrum}, AQUA~\cite{aqua}, Azure Data Lake Storage query acceleration~\cite{azure-accel}). Systems like Presto~\cite{presto}, PushdownDB~\cite{pushdowndb}, and FlexPushdownDB~\cite{fpdb} support pushdown via S3 Select, and FlexPushdownDB further hybrids pushdown with caching which benefits from both techniques.
% \aia{Do we need to talk more about FlexPushdownDB, since it is our own prior work?}
Recently, PolarDB-X~\cite{polardbx} has expanded its range of pushdown operators to include sorting and co-located equi-joins (both joining relations adopt the same partition function over the join key).

\subsection{Limitations of Computation Pushdown in Existing Cloud OLAP DBMSs} \label{sec:back-limit}

Even though computation pushdown is widely embraced by existing cloud OLAP DBMSs, there are still restrictions and unexplored opportunities. In the rest of this section, we demonstrate two major limitations to motivate the work in this paper.

\spara{Limitation 1: Static Pushdown Decisions Made at Planning Time} Existing cloud OLAP DBMSs make pushdown decisions during the query optimization phase --- the query plan is split into two pieces, with the pushdown portion executed in the storage layer, and the rest executed in the computation layer. Once a pushdown decision is made, it cannot be changed at runtime --- a pushdown task will always be executed at storage. In other words, pushdown decisions are made \textit{eagerly}, where the entire pushable subquery plan is offloaded to the storage.

\begin{figure}[h]
\begin{minipage}{0.55\linewidth}
    \centering
    \includegraphics[width=\linewidth]{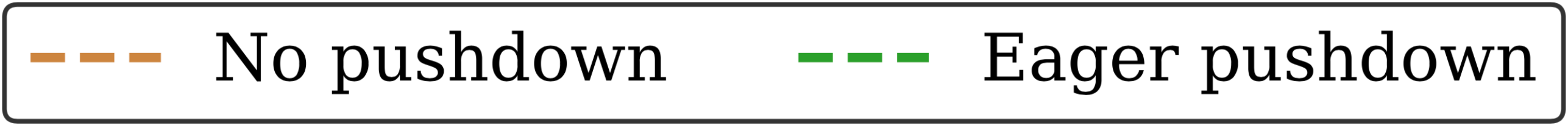}
\end{minipage}\\
\vspace{-0.15in}
\hspace{-0.3cm}
\begin{minipage}{\linewidth}
    \subfloat[Q1]{
        \includegraphics[width=0.5\linewidth]{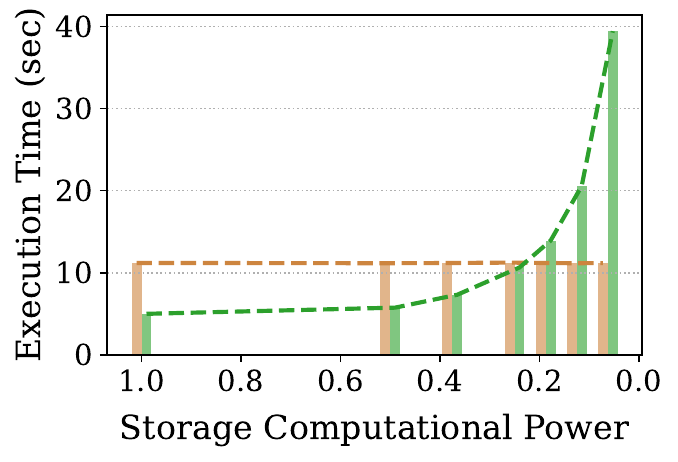}
    }
    \subfloat[Q19]{
        \includegraphics[width=0.5\linewidth]{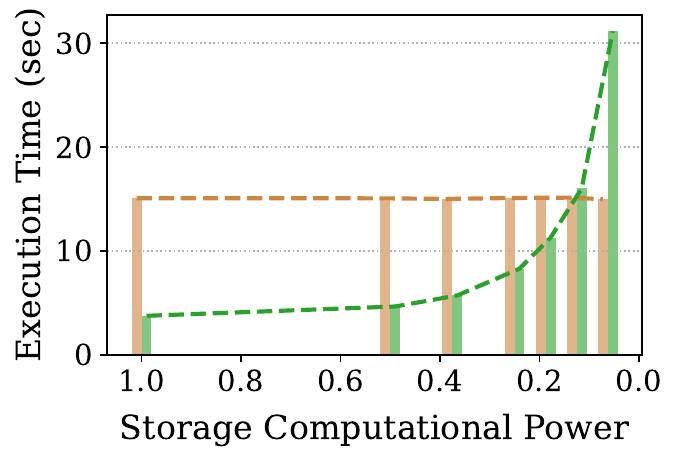}
    }
\end{minipage}
\vspace{-0.2cm}
\caption{Performance of \textit{No pushdown} and \textit{Eager pushdown} on Sample Queries --- \normalfont{Q1 and Q19 in TPC-H.}}
\label{fig:mot-adapt}
\end{figure}

Eagerly pushing all computation tasks that can be offloaded may not always benefit query processing, since the storage layer is in a shared multi-tenant environment, where the amount of computation resource for each request may vary. To build a deeper understanding, we prototype a S3-like object storage layer within \name, and measure the performance of pushdown using standard benchmark queries (TPC-H~\cite{tpch}, see Section~\ref{sec:eval-setup} for detailed setups) in different storage-layer resource utilization conditions. Figure~\ref{fig:mot-adapt} presents the results of two sample queries (Q1 and Q19). \np is included as a baseline for comparison.

\vspace{0.15cm}

For both queries, the performance of \np is independent of the available storage-layer computational power 
% \msc{define this metric}
(we regard full storage-layer computational power as all CPU cores at the storage node are available for pushdown tasks, see Section~\ref{sec:eval-adapt}). \ep outperforms \np when the storage has abundant computational resource, since it offloads computation tasks to the storage layer which reduces data transfer. However, when storage-layer computational resource is insufficient, query execution starts to suffer from the slowdown of pushdown execution. When the storage layer is saturated, \ep underperforms \np, and pushdown execution becomes the major performance bottleneck in the end-to-end query processing.

An ideal solution should consider the storage-layer computational resource utilization status, and adjust pushdown decisions at runtime \textit{adaptively}. Intuitively, when the storage system is idle or under light load, more computation tasks should be placed and executed at the storage layer to speed up query processing, which behaves close to \ep. Conversely, when the storage system is under heavy load, the DBMS should be more inclined to execute operators at the local compute nodes instead of offloading them to the storage, to avoid throttling storage-layer computation, which leads to a behavior similar to \np. We will present the proposed adaptive pushdown mechanism in Section~\ref{sec:adapt}.

\spara{Limitation 2: Lack of Systematic Analysis on Pushdown Operators} Existing systems design and implement pushdown features empirically, which end up picking a customized set of pushdown operators respectively. For example, simple pushdown functionalities like selection, projection, and scalar aggregation are supported by almost all existing pushdown systems, grouped aggregation is favored by Redshift Spectrum~\cite{spectrum}, and pushdown of bloom filters is introduced in PushdownDB~\cite{pushdowndb}.

We aim to conduct a comprehensive analysis of the design space, to identify key characteristics that contribute to the suitability of a pushdown operator. We closely examine the behaviors of existing OLAP DBMSs that offer pushdown support and categorize pushdown operators based on their key features. By deriving a shared pattern from these observations, we can establish a general principle that will potentially facilitate the discovery of new pushdown operators (Section~\ref{sec:op}).

\section{Adaptive Pushdown} \label{sec:adapt}

With computation pushdown, typically a query plan is split into two parts: the pushdown portion is executed in the storage layer, and the remaining part is executed in the computation layer. Existing cloud OLAP DBMSs execute the query plan with pushdown \textit{eagerly}, where the entire pushdown portion of the query plan is placed and executed in storage. However, as discussed in Section~\ref{sec:back-limit}, this may be sub-optimal since the storage layer may have insufficient computational resource at runtime (e.g., due to other pushdown requests), leading to performance degradation.

\vspace{0.15cm}
\begin{figure}[ht]
    \centering
    \includegraphics[width=0.9\linewidth]{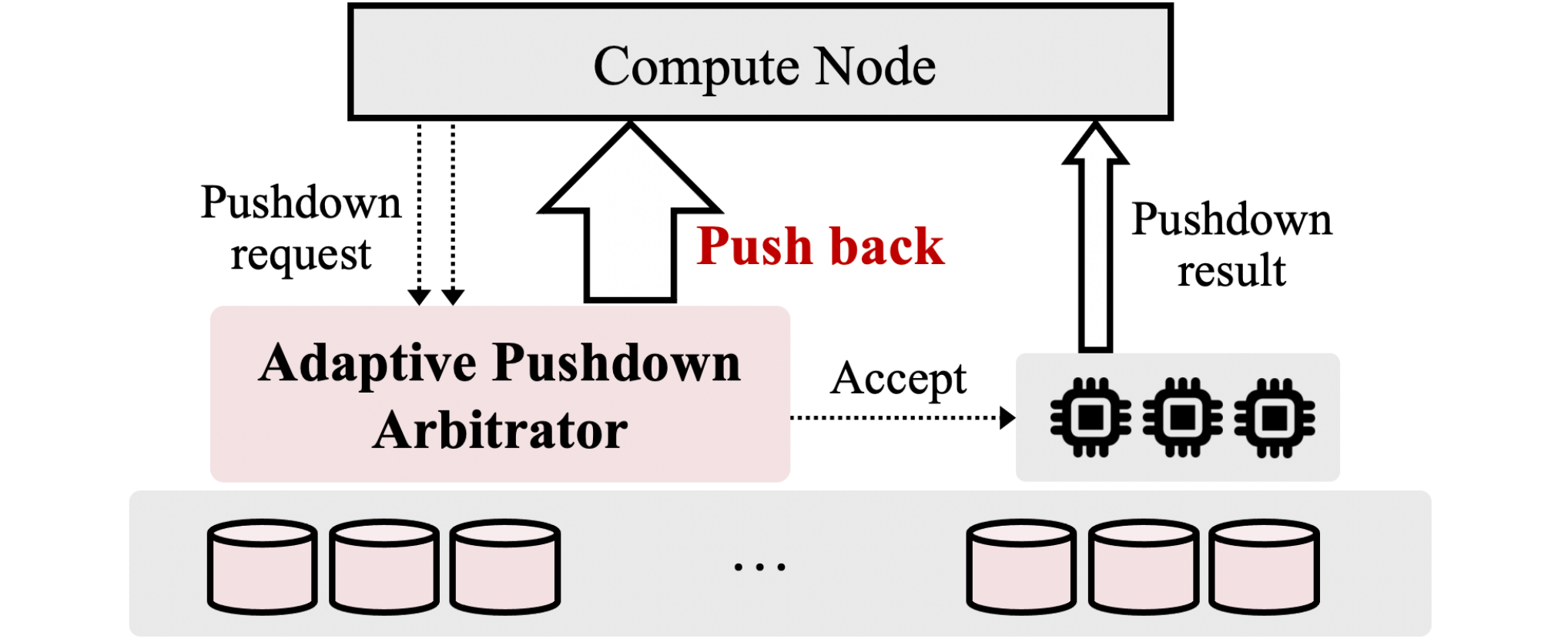}
    \vspace{-0.15cm}
    \caption{\textbf{Architecture of Adaptive pushdown --- \textnormal{The \textit{Adaptive Pushdown Arbitrator} determines whether to accept a pushdown request for execution or push it back.}}}
    \label{fig:adapt-arch}
\end{figure}

We propose that the DBMS should perform computation pushdown in storage \textit{adaptively}, by taking into account the storage-layer resource utilization status. The high-level architecture of adaptive pushdown is shown in Figure~\ref{fig:adapt-arch}. Instead of blindly executing all pushdown tasks in the storage layer, only a portion of tasks that the pushdown engine can sustain are offloaded. For the rest computation tasks, only the accessed raw data is returned to the compute nodes and no pushdown occurs. The two portions of computation results are then combined at the computation layer.

One major challenge of adaptive pushdown is that it is hard for the compute nodes to collect accurate statistics from the storage layer to decide whether a pushable task should be offloaded. Furthermore, even if such information can be accurately collected at planning time, the resource utilization in the storage layer may change at runtime. In this section, we develop a \textit{pushback} mechanism to let the storage layer instead of the computation layer to make the pushdown decision at runtime.

\subsection{Theoretical Analysis} \label{sec:adapt-theo}

We begin by analyzing the theoretical optimal bound of adaptive pushdown --- the optimal division of the computation tasks between pushdown and non-pushdown to achieve the best overall performance. Specifically, we analyze for the case of \textbf{a single query} under the following assumptions. 

% \msc{Note: I would not say ``for simplicity". You would get the optimal query execution time when a query is executed in isolation, without other concurrent queries. That's why you consider a single query. Is that correct?}

\begin{itemize}
    \item The query contains $N$ pushdown requests submitted to the storage layer in parallel. 
    % $N$ data partitions in the storage respectively.
    % \msc{Why do we need to talk about data partitions? We have not introduced this concept before and it does not seem to appear explicitly in the analysis. We are only assuming that the storage layer decides for each request, independently, whether to execute it or push it back.} % (e.g. to $N$ data partitions).
    % \item \msc{You are considering a single storage server right? Because otherwise you would have to consider how requests are distributed across servers, each having their resources.}
    \item Each pushdown request consumes the same amount of computational resource \texttwin{bw}{pd} when admitted, or the same amount of network resource \texttwin{bw}{pb} when pushed back. The total available CPU and network bandwidth at storage are \texttwin{BW}{cpu} and \texttwin{BW}{net} respectively.
    \item The overall execution time of the pushable query plan portion is \texttwin{T}{pd} with pushdown enabled, and \texttwin{T}{npd} with pushdown disabled, where $\frac{\eqtwin{T}{npd}}{\eqtwin{T}{pd}} = k$. Intuitively, $k$ determines the maximum speedup that any pushdown technique can possibly achieve.
\end{itemize}

We use the following terms to describe the pushdown decisions made at the storage layer.

\begin{itemize}
    \item Among $N$ pushdown requests arriving at the storage, $n$ requests are admitted and $N-n$ requests are pushed back.
    \item The admitted pushdown requests result in an overall execution time of $\eqtwin{T}{pd\_part}$, and requests that are pushed back lead to an overall execution time of $\eqtwin{T}{pb\_part}$. %\yxy{maybe call them $\eqtwin{T}{total\_pd}$ and $\eqtwin{T}{total\_pb}$ instead? }
\end{itemize}

Since the admitted pushdown requests at the storage layer and the pushback requests at the computation layer are executed in parallel, the overall execution time of the pushable portion in the query plan can be formulated as follows (Equation~\ref{eq:theo-overall}).
% \aia{This should be max not min.}

\vspace{-0.12cm}
\begin{equation}
\small
  T = max\{\eqtwin{T}{pd\_part}\algsp\algsp, \algsp\eqtwin{T}{pb\_part}\}
  \label{eq:theo-overall}
\end{equation}

In the optimal case, the storage layer would have a global view of all requests that it will receive ahead of the query execution. Therefore, an optimal split of the pushdown requests for admission and pushback can be constructed. Intuitively, with more pushdown requests admitted at the storage, $\eqtwin{T}{pd\_part}$ becomes larger, and $\eqtwin{T}{pb\_part}$ gets smaller, and vice versa. The overall execution time obtains the minimum when these two terms are equal, namely:
% \aia{The analysis above and the equation below need some more explanation. Consider $N$ queries of the form ``select x from T where x=const'' and assume x is unique so that the $N$ queries each return one row. As long as the storage layer has capacity, it is advantageous to push down as many of the $N$ queries as possible. Pushing a query grows $\eqtwin{T}{pd\_part}$ a little and shrinks $\eqtwin{T}{pb\_part}$ a lot.}

\vspace{-0.12cm}
\begin{equation}
\small
  \eqtwin{T}{opt} = \eqtwin{T}{pd\_part} = \eqtwin{T}{pb\_part}
  \label{eq:theo-opt-1}
\end{equation}

We assume the pushdown tasks are bounded by CPU computation and pushback tasks are bounded by network. Therefore, Equation~\ref{eq:theo-opt-1} can be further expanded as follows (Equation~\ref{eq:theo-opt-2}).

\vspace{-0.12cm}
\begin{equation}
\small
  \frac{n \cdot \eqtwin{bw}{pd}}{\eqtwin{BW}{cpu}} = \frac{(N-n) \cdot \eqtwin{bw}{pb}}{\eqtwin{BW}{net}}
  \label{eq:theo-opt-2}
\end{equation}

\texttwin{T}{pd} and \texttwin{T}{npd} can be expressed similarly (Equation~\ref{eq:theo-opt-3}).

\vspace{-0.12cm}
\begin{equation}
\small
  \eqtwin{T}{pd} = \frac{N \cdot \eqtwin{bw}{pd}}{\eqtwin{BW}{cpu}} , \hspace{2em} \eqtwin{T}{npd} = \frac{N \cdot \eqtwin{bw}{pb}}{\eqtwin{BW}{net}}
  \label{eq:theo-opt-3}
\end{equation}

Since we know $\frac{\eqtwin{T}{npd}}{\eqtwin{T}{pd}} = k$ by assumptions, we plug it into Equation~\ref{eq:theo-opt-3} and obtain Equation~\ref{eq:theo-opt-3'}.

\vspace{-0.12cm}
\begin{equation}
\small
  \frac{kN \cdot \eqtwin{bw}{pd}}{\eqtwin{BW}{cpu}} = \frac{N \cdot \eqtwin{bw}{pb}}{\eqtwin{BW}{net}}
  \label{eq:theo-opt-3'}
\end{equation}

Combining Equation~\ref{eq:theo-opt-3'} and Equation~\ref{eq:theo-opt-2}, we can express $n$ as follows (Equation~\ref{eq:theo-opt-5}).
% \aia{It is not clear which equation you are solving to get the expression below.} 

\vspace{-0.12cm}
\begin{equation}
\small
  n = \frac{k}{k+1}N
  \label{eq:theo-opt-5}
\end{equation}

Additionally, the optimal execution time can be expressed as follows (Equation~\ref{eq:theo-opt-6}).

\vspace{-0.12cm}
\begin{equation}
\small
  \eqtwin{T}{opt} = \frac{k}{k+1}\eqtwin{T}{pd} = \frac{1}{k+1}\eqtwin{T}{npd}
  \label{eq:theo-opt-6}
\end{equation}

Intuitively, a larger $k$ means a higher speedup through pushdown, leading to more tasks executed in the storage ($n$ being larger). Even when the pushdown layer processes tasks slower than the compute layer ($k < 1$), it can still accelerate the system by offloading some computation. When the pushdown layer is extremely slow or does not exist ($k = 0$), all tasks are pushed back to the compute layer and no pushdown occurs. 
Note the optimum (Equation~\ref{eq:theo-opt-5}) can only be approximately satisfied in practice, since the number of pushdown and pushback tasks can only be integers. For example, assume a query submits ten requests to storage, and the optimal division of pushdown and pushback tasks is 7.7 versus 2.3. In practice we have to round them to the closest integers, i.e., execute eight requests at the storage and push back the rest two requests.

% \yxy{I think a bit more discussion can be added here to help readers understand the last equation. larger k means higher speedup through pushdown, which means more tasks should be pushed down. Also, you may want to add an equation for T. like $T = T_{npd} * (k+1)$. Here, we also need to discuss this equation. For example, I think k=0 means pushdown as no gain, which corresponds to the case where pushdown layers is extremely slow or does not exist, so that $T_{pd}=\infty$. k<1 is also meaningful, which means the pushdown layer can process the query slower than the compute layer, but it can speedup the system by offloading some computation. Not sure my last sentence makes sense, but we need to somehow discuss it in the paper.}
\subsection{Pushback Mechanism} \label{sec:adapt-mech}

In our design, the compute nodes always try to offload all pushdown tasks as if the storage has abundant computational resource. When the storage server receives a pushdown request, the \textit{Adaptive Pushdown Arbitrator} (see Figure~\ref{fig:adapt-arch}) determines whether the pushdown request should be accepted and executed. If the storage server is busy, the pushdown request is rejected and the computation task is \textit{pushed back}, in which case the raw data is returned and processed at the compute node as if pushdown did not happen. The query plan at the computation layer is then dynamically adjusted to accommodate the pushback.

\begin{algorithm}[h]
\small
\DontPrintSemicolon
\SetKwInOut{State}{State}
\State{wait queue: \texttwin{Q}{wait}, pushdown execution slots: \texttwin{S}{exec-pd}\\
       pushback execution slots: \texttwin{S}{exec-pb}}
\SetKwInOut{Assume}{Assume}
\Assume{all incoming requests are first enqueued into \texttwin{Q}{wait}}
{
    \While{\texttwin{Q}{wait} \textnormal{is not empty}}{
        \textit{req} = \texttwin{Q}{wait}\algsp.front\algsp() \\
        \texttwin{t}{pd} = estimate\_pushdown\_time\algsp(\textit{req}) \\
        \texttwin{t}{pb} = estimate\_pushback\_time\algsp(\textit{req}) \\
        \eIf{\texttwin{t}{pd} < \texttwin{t}{pb}}{
            \textit{success} = try\_pushdown\algsp(\textit{req}, \texttwin{S}{exec-pd}) \textbf{or} \\
                \hspace{3.6em} try\_pushback\algsp(\textit{req}, \texttwin{S}{exec-pb})
        }{
            \textit{success} = try\_pushback\algsp(\textit{req}, \texttwin{S}{exec-pb}) \textbf{or} \\
                \hspace{3.6em} try\_pushdown\algsp(\textit{req}, \texttwin{S}{exec-pd})
        }
        \eIf{success}{
            \texttwin{Q}{wait}\algsp.dequeue\algsp() \\
        }{
            \textbf{break}
        }
    }
}
\caption{\small{Pushback Mechanism at the Storage Layer}}
\label{alg:mech}
\end{algorithm}

Algorithm~\ref{alg:mech} illustrates the pushback mechanism deployed at the storage layer. It is invoked when a new request arrives or a running request completes. The state maintained in the pushdown node includes a \textit{wait queue} (\texttwin{Q}{wait}), which is used to buffer excess pushdown requests when the server is under heavy load, and a finite set of \textit{execution slots} (\texttwin{S}{exec-pd}, \texttwin{S}{exec-pb}) for both pushdown and pushback executions, to help isolate performance among different executions and avoid throttling. We assume all incoming requests are first enqueued into \texttwin{Q}{wait}.
% , virtualized \yxy{the meaning of the word "virtualized" is a bit unclear} from the computation and network resources of the storage server respectively. The virtualization helps isolate performance among different executions and avoid throttling.

For each request in the wait queue (line 1--2), we begin by estimating the execution time for both pushdown and pushback (lines 3--4), which are classified as the faster path and slower path respectively through comparison (line 5). The details on the time estimations will be discussed in Section~\ref{sec:adapt-model}. 
The Adaptive Pushdown Arbitrator first tries to assign the request to the faster path (line 6 and line 9). If the assignment is not successful due to resource contention --- the corresponding execution slots are full, the Arbitrator then tries to assign the request to the slower path (line 7 and line 10). If at least one assignment is successful, the request is removed from the wait queue and executed correspondingly, and we start evaluating the next request in the wait queue. Otherwise, the process stops since currently both computation and network resources are saturated. The intuition here is that the storage server should balance the resource utilization between CPU and network at runtime adaptively.
% If neither assignment is successful --- both computation and network resources are saturated, the request is placed into the wait queue (line 8 and line 14). \aia{Which part of the system will de-queue the enqueued requests, and when will this happen? This is not clear here and it is not clear  later when you talk about pushdown amenable operators.} 

% \spara{Double Execution for Tail Requests} For a single query, the executions on the pushdown path and pushback path may not finish at the same time. For instance, when all the pushdown executions have finished, a few tail pushback tasks may still be running and taking relatively long time to finish. In this case, it would have been better if some of these pushback tasks are instead executed through pushdown. %, which may block the downstream operators and hurt the performance. 
% The phenomenon typically occurs when the speedup of pushdown ($k$) is extremely high, where a single pushback execution may run slower than all the other pushdown executions. We develop the \textit{double execution} mechanism to alleviate the issue. For each tail running request (e.g., pushback), we start a mirror execution in the opposite path (e.g., pushdown). The compute layer receives the output of whichever finishes earlier, and ignores the slower execution. Note that double execution may consume extra resources in the storage layer, so it is triggered only when no more requests are arriving, such that the storage server is idle.
\subsection{Lightweight Model for Time Estimation} \label{sec:adapt-model}

There are potentially many ways to estimate the execution time of pushdown and pushback aforementioned. For simplicity, we choose a straightforward formulation to represent them by the estimated total amount of work (measured in time).

\spara{Pushdown Time} The pushdown execution consists of three components: data scanning, computation, and network transfer of pushdown results (Equation~\ref{eq:pd}).

\vspace{-0.12cm}
\begin{equation}
\small
  \eqtwin{t}{pd} = \eqtwin{t}{scan} + \eqtwin{t}{compute} + \eqtwin{t}{net}
  \label{eq:pd}
\end{equation}

\texttwin{t}{compute} and \texttwin{t}{net} can be estimated as follows (Equations~\ref{eq:pd-comp-net}).

\vspace{-0.12cm}
\begin{equation}
\small
  \eqtwin{t}{compute} = \frac{\eqtwin{S}{in}}{\eqtwin{C}{storage}}, \hspace{2em} \eqtwin{t}{net} = \frac{\eqtwin{S}{out}}{\eqtwin{BW}{net}}
  \label{eq:pd-comp-net}
\end{equation}

\texttwin{S}{in} and \texttwin{S}{out} denote the size of accessed data and pushdown results. For row-oriented formats (e.g., CSV), \texttwin{S}{in} is the size of the input data object. For column-oriented formats (e.g., Parquet), \texttwin{S}{in} is the size of all accessed columns. \texttwin{S}{out} is subject to the selectivity of the pushdown operation, which can be estimated using cardinality estimation techniques~\cite{kim2022learned, harmouch2019cardinality}. \texttwin{C}{storage} refers to the computation bandwidth at storage, depending on the type of the pushdown operation and the number of requested CPU cores. \texttwin{t}{compute} can be computed by either estimating \texttwin{C}{storage} by performing micro-benchmarks on different operators at the storage servers, or utilizing existing techniques of execution time estimation~\cite{chaudhuri2004estimating, lee2016operator, wu2013towards, luo2004toward, duggan2011performance}. As assumed in Section~\ref{sec:adapt-mech}, a fixed amount of network bandwidth is dedicated to each pushdown request, so \texttwin{BW}{net} is a known constant.
% \msc{How do you ensure this in practice? How do you dedicate a fraction of bandwidth to each requests?}

\spara{Pushback Time} The execution time for a request that is pushed back to the computation layer can be modeled by data scanning and raw data transfer. Here we ignore the computation in the compute layer, since in a storage-disaggregated architecture, usually raw data transfer dominates the pushback time so the computation component has little effect (in Section~\ref{sec:op-prin} we observe most existing pushdown operators are bounded). 
% \msc{This actually depends on the operator. It would follow by your assumption that pushdown operators are bounded.}
Besides, the storage layer is unaware of the computation bandwidth of the compute nodes, which can vary across different users.

\vspace{-0.12cm}
\begin{equation}
\small
  \eqtwin{t}{pb} = \eqtwin{t}{scan} + \eqtwin{t'}{net}
  \label{eq:pb}
\end{equation}

\texttwin{t'}{net} can be estimated as follows (Equation~\ref{eq:pb-net}).

\vspace{-0.12cm}
\begin{equation}
\small
  \eqtwin{t'}{net} = \frac{\eqtwin{S}{in}}{\eqtwin{BW}{net}}
  \label{eq:pb-net}
\end{equation}

Finally, note that a detailed analysis on \texttwin{t}{scan} (time to scan data from disks in the storage) is unnecessary, since it is included in both pushdown time and pushback time, and will cancel each other when compared in Algorithm~\ref{alg:mech} (line 5).
\subsection{Awareness of Pushdown Amenability} \label{sec:adapt-pa}

The wait queue deployed in Algorithm~\ref{alg:mech} manages the arriving pushdown requests in a FIFO order, which treats all the requests identical. However, some requests benefit more on pushdown compared to other requests (e.g., the request has a selective filter but incurs little computation). Intuitively, these pushdown-amenable requests should be given a higher priority to be executed in the pushdown path than the requests that cannot benefit a lot by pushdown.

Consider a scenario where the wait queue contains two requests: $r_1$ with $\eqtwin{t}{pd} = 3$ and $\eqtwin{t}{pb} = 4$, followed by $r_2$ with $\eqtwin{t'}{pd} = 1$ and $\eqtwin{t'}{pb} = 4$. The two requests have the same estimated pushback time but differs on the estimated pushdown time. Assume at a moment one request can be admitted for pushdown execution and the other needs to be pushed back. Algorithm~\ref{alg:mech} would first evaluate $r_1$ and places it into the pushdown path, then evaluate $r_2$ with it pushed back. However, a better solution would be to push back $r_1$ instead of $r_2$, since $r_2$ incurs a lower execution time by pushdown.

Given a request, we define \textit{Pushdown Amenability} as the potential benefit of pushdown compared to pushback, which can be expressed as follows (Equation~\ref{eq:pa}).

\vspace{-0.12cm}
\begin{equation}
\small
  PA = \eqtwin{t}{pb} - \eqtwin{t}{pd}
  \label{eq:pa}
\end{equation}

At runtime, the Adaptive Pushdown Arbitrator keeps the wait queue sorted by the \textit{PA} value of the requests. Pushdown execution always consumes the request with the highest \textit{PA} value, and pushback execution does the reverse. In the example shown above, $PA(r_1) = 1$ and $PA(r_2) = 3$, such that the storage server would consider executing $r_2$ in the storage and pushes $r_1$ back. Compared to Algorithm~\ref{alg:mech}, the invariant here is the full utilization of both the computational and network resources. However, the total amount of consumed CPU and network resources are potentially decreased.

\section{Analysis of Pushdown Operators} \label{sec:op}

In this section, we aim to obtain a comprehensive understanding of the effects of computation pushdown to cloud OLAP DBMSs. We collect the supported pushdown operators in existing cloud DBMSs, and by observing the common patterns, we derive a general principle about whether an operator is amenable to pushdown (Section~\ref{sec:op-prin}). Following the principle, we further identify two operators that can benefit from pushdown to the storage, in particular, \textit{selection bitmap} and \textit{distributed data shuffle} (Section~\ref{sec:op-propose}).

\subsection{Key Characteristics} \label{sec:op-prin}

\begin{table*}[t]% h asks to places the floating element [h]ere.
  \caption{Supported Pushdown Operators in Existing Cloud OLAP DBMSs --- \textnormal{Systems supporting pushdown via S3 Select with no additional enhancement are represented as \textit{S3 Select} (e.g., Presto). Marked with *: Pushdown of grouped aggregation and bloom filters are not efficiently supported by PushdownDB, join pushdown in PolarDB-X requires both tables co-located on the join key, and top-K pushdown in PolarDB-MySQL requires indexes on the sort key.}}
  \vspace{-0.2cm}
  \label{tab:op-exist}
  \footnotesize
  \begin{tabular}{cccccccccc}
    \toprule
    \textbf{Operator} & Selection & Projection & Scalar Aggregation & Grouped Aggregation & Bloom Filter & Top-K & Sort & Join & Merge\\
    \cmidrule{1-10}\morecmidrules\cmidrule{1-10}
    \textbf{Redshift Spectrum} & \checkmark & \checkmark & \checkmark & \checkmark\\
    \cmidrule{1-10}
    \textbf{AQUA} & \checkmark & \checkmark & \checkmark\\
    \cmidrule{1-10}
    \textbf{S3 Select} & \checkmark & \checkmark & \checkmark\\
    \cmidrule{1-10}
    \textbf{PushdownDB} & \checkmark & \checkmark & \checkmark & \checkmark$^*$ & \checkmark$^*$ & \checkmark &\\
    \cmidrule{1-10}
    \textbf{\begin{tabular}{@{}c@{}}Azure Data Lake\\ Query Acceleration\end{tabular}} & \checkmark & \checkmark & \checkmark\\
    \cmidrule{1-10}
    \textbf{PolarDB-X} & \checkmark & \checkmark & \checkmark & \checkmark & & \checkmark & \checkmark & \checkmark$^*$\\
    \cmidrule{1-10}
    \textbf{PolarDB-MySQL} & \checkmark & \checkmark & & & \checkmark & \checkmark$^*$ &\\
    \bottomrule
  \end{tabular}
\end{table*}

Computation pushdown in existing cloud OLAP DBMSs generally target intuitive pushdown operations. For example, selection, projection, and aggregation are mostly considered due to the ease of design and development, as presented in Table~\ref{tab:op-exist}. 
% \yxy{this sentence has redundant content that was mentioned in the previous paragraph}
Besides these simple pushdown functionalities, different systems adopt a customized set of pushdown operators. For instance, pushdown of grouped aggregation is incorporated by Redshift Spectrum. PushdownDB supports pushdown of grouped aggregation, top-K, and bloom filters using existing APIs of S3 Select. However, due to the intrinsic restrictions of S3 Select interfaces, the implementations are not as efficient as they could be. For example, pushdown of grouped aggregation has to be processed in two phases, resulting two rounds of data exchange. Besides, bloom filters are required to be serialized explicitly into strings with 0s and 1s, which is space- and computation-inefficient. Moreover, PolarDB-X enables pushdown for a specific type of join, that requires both tables to be co-partitioned on the join key.

We conclude the following key characteristics from Table~\ref{tab:op-exist} that contribute to the suitability of a pushdown operator.

\begin{tcolorbox}[boxsep=1pt, arc=1pt, boxrule=1pt, left=5pt, right=5pt, top=1pt, bottom=5pt]
    \spara{Key Characteristics of Pushdown} \textit{The required storage-layer computation is \uline{local} and \uline{bounded}.}
\end{tcolorbox}

\spara{Characteristic 1: Locality} Locality means the computation tasks placed at the storage layer do not incur any network traffic across the storage servers --- the traffic occurs only between the storage layer and the computation layer.

\sspara{Analysis of Operators} Simple pushdown logics like selection, projection, and scalar aggregation comply with locality, since no network traffic is incurred across storage nodes. The same rationale also applies to operators including grouped aggregation, bloom filter, top-K, sort, where the computation functionality can be performed on each individual data object.

The general join does not embrace locality, unless the % since how it is performed depends on the distribution of the two joining relations. When the 
two joining relations are co-partitioned using the join key. Otherwise the data needs to be shuffled across the network to redistribute the tables. %, a broadcast step on the smaller relation or a redistribution step on both relations are required before the join is performed, which incurs network traffic across the storage servers. However, when both relations are co-located on the join key, the join operation owns locality. 
PolarDB-X supports pushdown of co-partitioned joins. 
Another example of non-local operator is \textit{Merge}, which combines the output of multiple upstream operators (e.g., select, project, aggregate, sort, etc.). %is performed when a database table is partitioned into multiple shards at storage, where the results of a specific operation over all the partitions must be merged to produce the final output (e.g., select, project, aggregate, sort, etc.). 
Merge requires data exchange within the storage layer since data objects are typically spread across multiple storage severs, making it a non-local operation. As Table~\ref{tab:op-exist} shows, none of the existing cloud DBMSs supports pushing merge to the storage.

\sspara{Potential Advantages} We believe locality is an important characteristic for pushdown operators for three reasons. First, a pushdown environment must support multi-tenancy. Forbidding data exchange across storage servers can reduce the performance variations in pushdown tasks (e.g., due to network interference and queueing). 
%is critical in in a multi-tenant environment, 
%it is crucial to achieve isolation of performance and resource allocation between tenants. Pushdown computation without any data transfer across the storage nodes seldom incurs high performance variations that are hard to predict --- an unexpected data transfer between two storage nodes can hurt performance of regular data access in both nodes.
Second, cloud storage must be encrypted using %cloud storage vendors tend to utilize cryptographic 
protocols like TLS~\cite{tls} during data transfer. % to prevent data leaking. 
Local operators avoid the complexity of encryption and decryption across storage servers (e.g., distributing private keys). 
Third, local pushdown keeps the design of the storage layer simple, since it needs to support the server-side functionalities of client-server APIs.
%processes can incur considerable overhead and also increase system complexity, like sharing the private key across multiple storage servers. Pushdown operators with locality eliminate such additional entanglement.
%Finally, unlocking data transfer across storage servers for pushdown computation poses challenges to data migration of the storage layer, since it brings in additional network overhead, which can interfere with the data migration process and affect system availability. It is also non-trivial to determine the destination servers of data transfer prior to the execution of pushdown tasks, since the migration or replication process may change the placement of data objects, which can only be addressed at runtime.

\spara{Characteristic 2: Boundedness} \textit{Bounded} implies that pushdown tasks should only require linear amount of CPU and memory resources with regard to the accessed data size.

\sspara{Analysis of Operators} Selection, projection, and scalar aggregation are linearly bounded since the CPU consumption is linear to the size of the processed data, and the memory consumption is a constant. Grouped aggregation consumes linear CPU and linear memory capacity. %is typically hash-based on unsorted data. For data sorted on the group key, grouping can be done in a single pass through the data. Both methods consumes linear amount of CPU and memory resource. 
Bloom filter can be regarded as a special regular filter which is thus also linearly bounded. Top-K is typically implemented using a max or min heap which consumes $O(K)$ memory and $O(NlogK)$ execution time, where $N$ is the size of the input data.  The variable $K$ is a constant and typically much smaller than $N$, making the time complexity also linear in $N$.

The computation complexity of the sort operation is not linearly bounded. % beyond linearity, and thus is not linearly bounded. 
The boundedness of the join operator depends on the cardinality of its output. In a key-foreign key join, the output size is bounded by the larger table, which also serves as the upper bound for both memory usage and computational complexity. However, if the join is not an equi-join, it needs be computed using a nested loop, resulting in complexity that grows beyond linear. Only PolarDB-X incorporates join pushdown for co-partitioned tables. %them with restrictions.

\sspara{Potential Advantages} Supporting only \textit{bounded} operators preserves the key benefits of storage-disaggregation, where the storage service scales only %. The principle of storage-disaggregation is to scale the number of storage servers only
based on the volume of the stored data, regardless of the computational resource consumption of the workloads. If pushdown operators are not bounded (linearly), the computational resource consumption may grow beyond the scale of the store data, which requires the storage servers to balance between storage and computational needs, which defeats the purpose of storage-disaggregation. 
%In this case, the system behaves more similar to a shared-nothing architecture, which sacrifices desirable properties including elastic resource management and separation of concerns. When pushdown computation is bounded, the benefits of disaggregation can be preserved because complex and expensive operators are still evaluated at the computation layer.

\subsection{Proposed Pushdown Operators} \label{sec:op-propose}

Following the key characteristics derived from existing systems, we identify two new operators that can also benefit from pushdown to storage, but have not been deeply investigated previously.

\spara{Selection Bitmap} Late materialization is widely adopted by columnar OLAP engines, as demonstrated by various previous studies~\cite{stonebraker2005cstore, polychroniou2014track, lin2011llama, abadi2008column, raman2013db2}. Selection bitmaps are one common technique embracing late materialization, and DBMSs frequently filter columnar data using selection bitmaps. For instance, when selecting a column based on a filter predicate on another column, the predicate column is read in first to generate a selection bitmap. This bitmap is then used to filter the selection column. Similar ideas appear in joins, where the join columns will generate a bitmap, which is used to select the non-join columns.

However, with storage-disaggregation becoming more prevalent, selection bitmaps and the columns they filter are often situated on opposite ends of the network. For example, the column to be selected may reside in the local cache of compute nodes, while the predicate column is stored in the remote cloud storage. Traditionally, it is necessary to transfer either column over the network to process the filter. A better solutions is to ship the selection bitmap instead, which can potentially result in much less network traffic.

\sspara{Selection Bitmap from the Storage Layer} Selection bitmaps can be transferred from the storage to the computation layer when they can only be created at the storage layer. We use the query below as an example to demonstrate how selection bitmap pushdown works. The query essentially evaluates a filter predicate on attribute $B$ and returns both columns $A$ and $B$.

\lstset{basicstyle=\ttfamily\small}
\begin{lstlisting}[caption=An Example Filtering Query.,label=lst:example,captionpos=b]
               SELECT A, B FROM R
               WHERE B > 10
\end{lstlisting}
\vspace{-0.1cm}

\begin{figure}[h]
	\centering
    \hspace{-0.2cm}
    \subfloat[Conventional]{
        \includegraphics[height=1.8in]{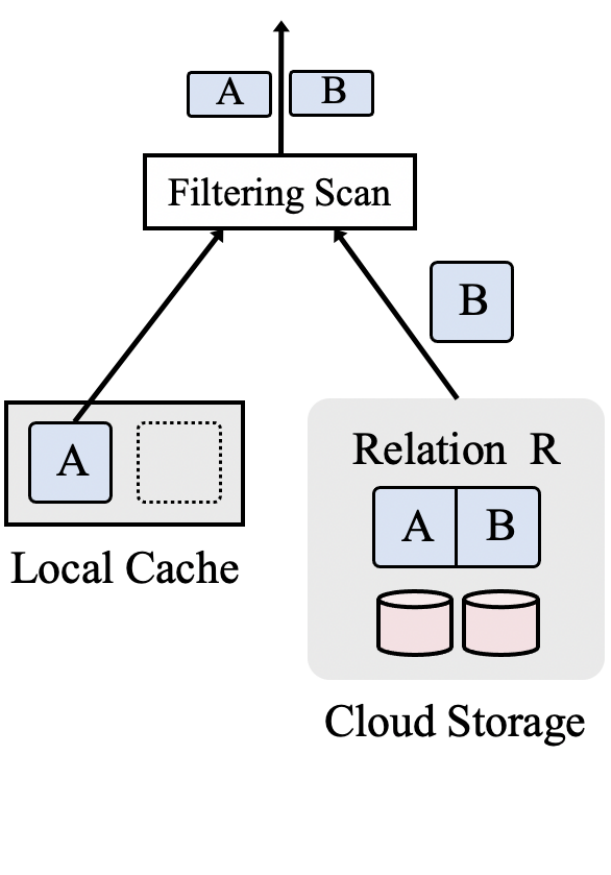}
    } 
	\hspace{0.3cm}
    \subfloat[With Selection Bitmap Pushdown]{
        \includegraphics[height=1.8in]{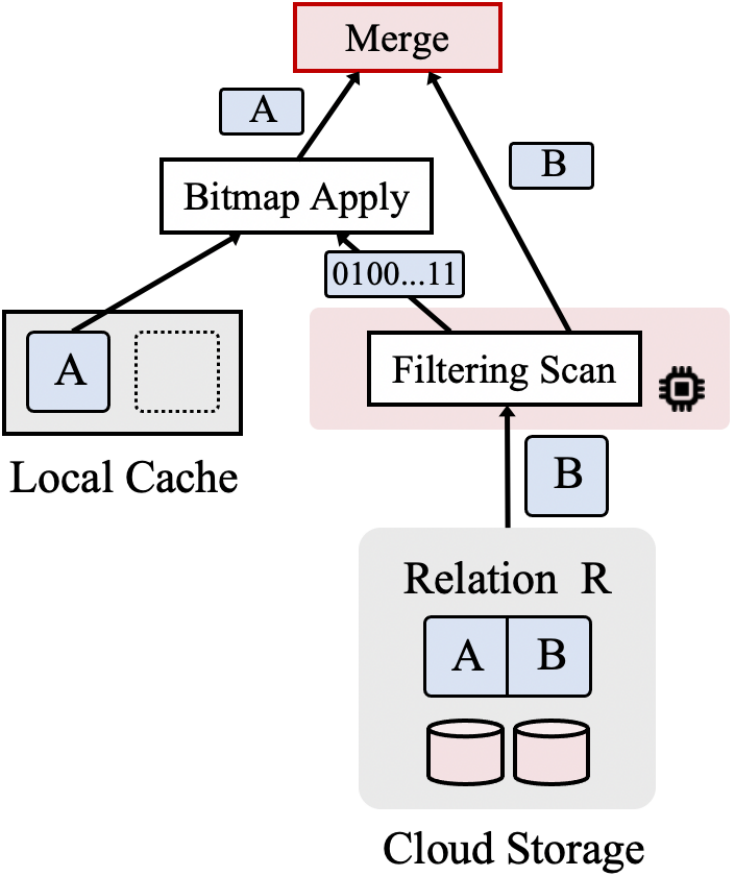}                
    }
    \vspace{-0.2cm}
    \caption{\textbf{Selection Bitmap Pushdown (from the Storage Layer) --- \textnormal{The selection bitmap constructed at storage can be used to filter cached data at the computation layer.}}}
    \label{fig:bitmap-pd-st}
\end{figure}

Assume column $A$ is stored in the local cache, as Figure~\ref{fig:bitmap-pd-st} shows. 
% \aia{The paper is very silent about caching. Should you elaborate on how caching interacts with pushdown?} 
Conventionally, the DBMS needs to load the missing column $B$ and evaluate the filter locally (Figure~\ref{fig:bitmap-pd-st}(a)) (it is also possible that the DBMS pushes down the entire scan and loads both filtered $A$ and $B$). The hybrid solution proposed in ~\cite{fpdb} also requires the predicate column $B$ loaded to the compute node.
Figure~\ref{fig:bitmap-pd-st}(b) illustrates the case when selection bitmap pushdown is enabled. The storage layer first constructs a selection bitmap by evaluating the filter predicate on column $B$. Then the bitmap is sent to the compute node, where column $A$ is loaded from the cache and filtered by applying the bitmap. At the same time, the filtered column $B$ is returned from storage. Finally, filtered $A$ and $B$ are merged at the compute node.

\begin{figure}[h]
	\centering
    \subfloat[Conventional]{
        \includegraphics[height=1.5in]{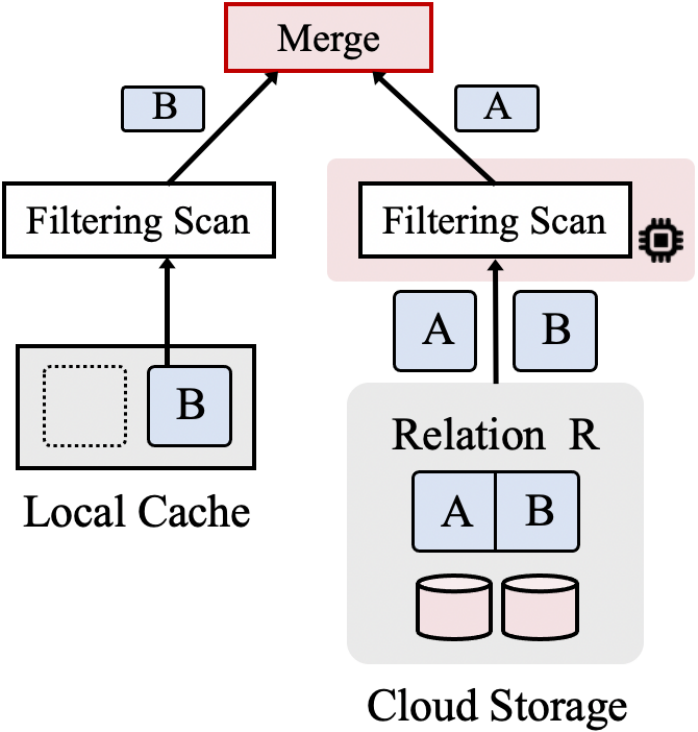}
    } 
	\hspace{0.1cm}
    \subfloat[With Selection Bitmap Pushdown]{
        \includegraphics[height=1.5in]{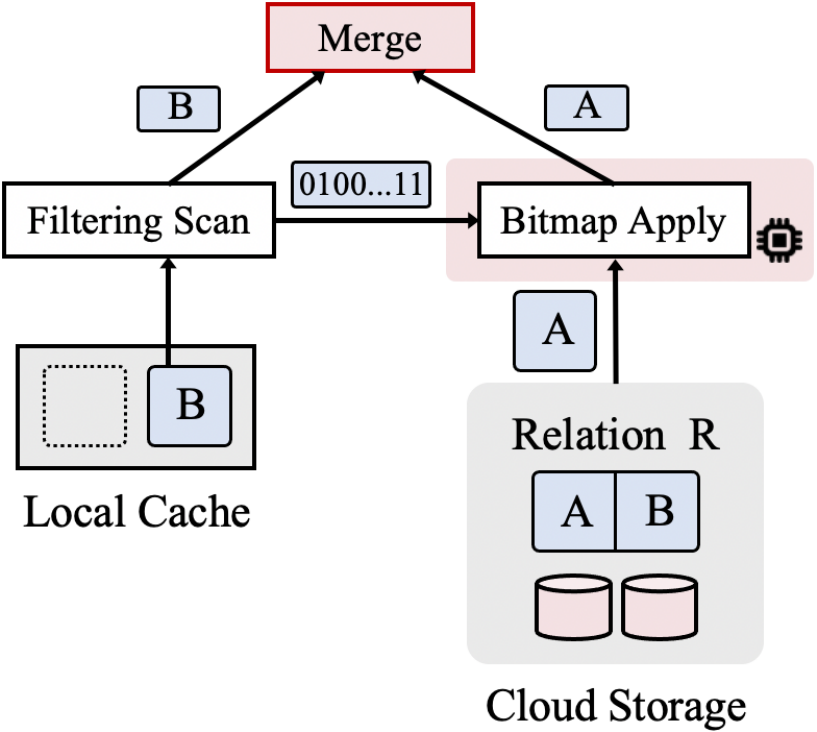}                
    }
    \vspace{-0.2cm}
    \caption{\textbf{Selection Bitmap Pushdown (from the Computation Layer) --- \textnormal{Storage can use the compute-layer selection bitmap to perform filtering without touching the predicate column.}}}
    \label{fig:bitmap-pd-cp}
\end{figure}

\sspara{Selection Bitmap from the Computation Layer} Similarly, the storage can utilize selection bitmaps constructed at the computation layer to perform filtering. We use the same example query as above (Listing~\ref{lst:example}) and assume column $B$ is cached instead, as Figure~\ref{fig:bitmap-pd-cp} shows. Conventionally, the storage layer must scan both columns $A$ and $B$ to evaluate the filter (Figure~\ref{fig:bitmap-pd-cp}(a)). With selection bitmap pushdown, as Figure~\ref{fig:bitmap-pd-cp}(b) demonstrates, the compute node can construct a selection bitmap when filtering $B$ from the local cache. The bitmap is then sent to storage such that the storage server can perform filtering without loading the predicate column $B$ from disks. Moreover, the CPU cycles used to compute the selection bitmap are eliminated at the storage layer.

Selection bitmap pushdown essentially is a variant of filtering pushdown, and hence is both local and bounded.

\sspara{Discussion on the Design Space} Selection bitmap pushdown from the storage layer and the computation layer complements each other, as one of them is always applicable depending on where the data columns locate. In practice, filter predicates are often composed of multiple sub-predicates connected by `AND' or `OR', and these sub-predicates may themselves be composite. In these situations, a fine-grained execution framework can be used to combine the benefits of both storage-side and compute-side selection bitmaps. For example, on evaluating the predicate ``($A$ or $B$) and $C$'', assume columns $A$ and $B$ are cached. We can assign the sub-predicate ``$A$ or $B$'' to be evaluated at the compute layer, while the rest ``$C$'' is left to storage. Both evaluations construct selection bitmaps and are exchanged, allowing for the formation of a complete selection bitmap that corresponds to the input filter predicate through inexpensive bitwise operations. Subsequently, the complete selection bitmap is used to filter both the cached columns at the computation layer and the uncached columns at the storage layer.

% \begin{figure}[h]
% 	\centering
%     \subfloat[Filter predicate]{
%         \includegraphics[height=1in]{figures/bitmap-fine-a.pdf}
%     } 
% 	\hspace{0.6cm}
%     \subfloat[Sub-tree Assignment]{
%         \includegraphics[height=1in]{figures/bitmap-fine-b.pdf}                
%     }
%     \caption{\textbf{Fine-grained Selection Bitmap Pushdown --- \textnormal{Sub-trees of the filter predicate are assigned according to the cached data.}}}
%     \label{fig:bitmap-pd-fine}
% \end{figure}

% A filter predicate can be represented as a tree, as illustrated in Figure~\ref{fig:bitmap-pd-fine}(a). The sub-trees of the filter predicate are assigned to either the computation layer or the storage layer for evaluation, depending on where the data columns locate. For instance, in Figure~\ref{fig:bitmap-pd-fine}(b), the sub-tree ``$A = 20$ OR $B = 20$'' is assigned to the compute node, while the sub-tree ``$C = 30$'' is left to storage. Selection bitmaps are constructed accordingly, based on sub-tree assignment. The storage-side and computation-side selection bitmaps are then exchanged, allowing for the formation of a complete selection bitmap that corresponds to the input filter predicate through inexpensive bitwise operations. Subsequently, the complete selection bitmap is used to filter both the cached columns at the computation layer and the uncached columns at the storage layer.

% need one empty line for spara

\begin{figure}[ht]% h asks to places the floating element [h]ere.
	\centering
    \hspace{-0.3cm}
    \subfloat[No Shuffle Pushdown]{
        \includegraphics[width=0.39\linewidth]{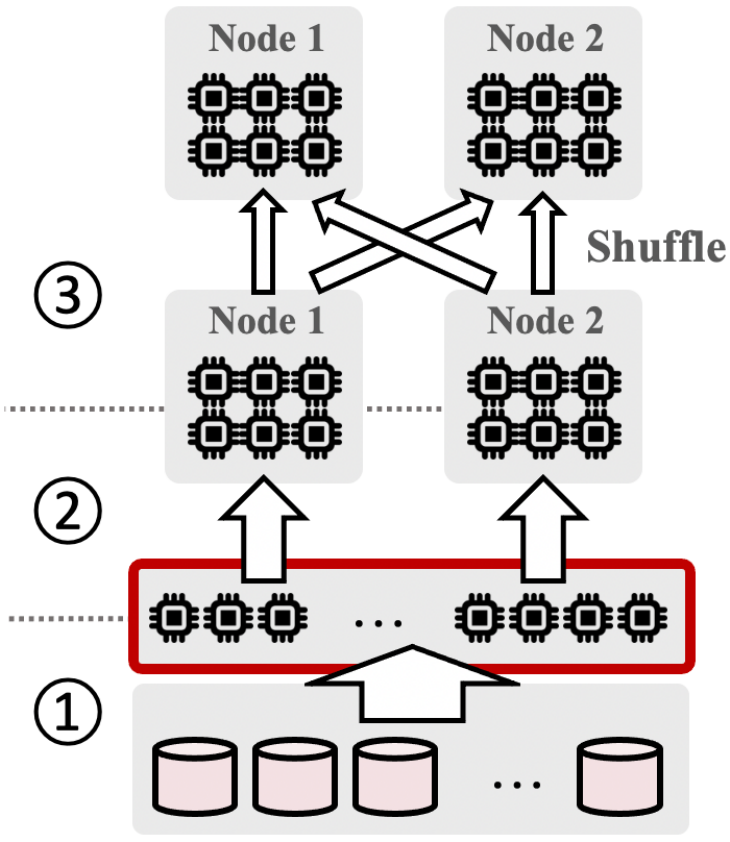}
    } 
    \hspace{-0.1cm}
	\captionsetup[subfigure]{singlelinecheck=off, margin={0.4cm,0cm}}
    \subfloat[With Shuffle Pushdown]{
        \includegraphics[width=0.6\linewidth]{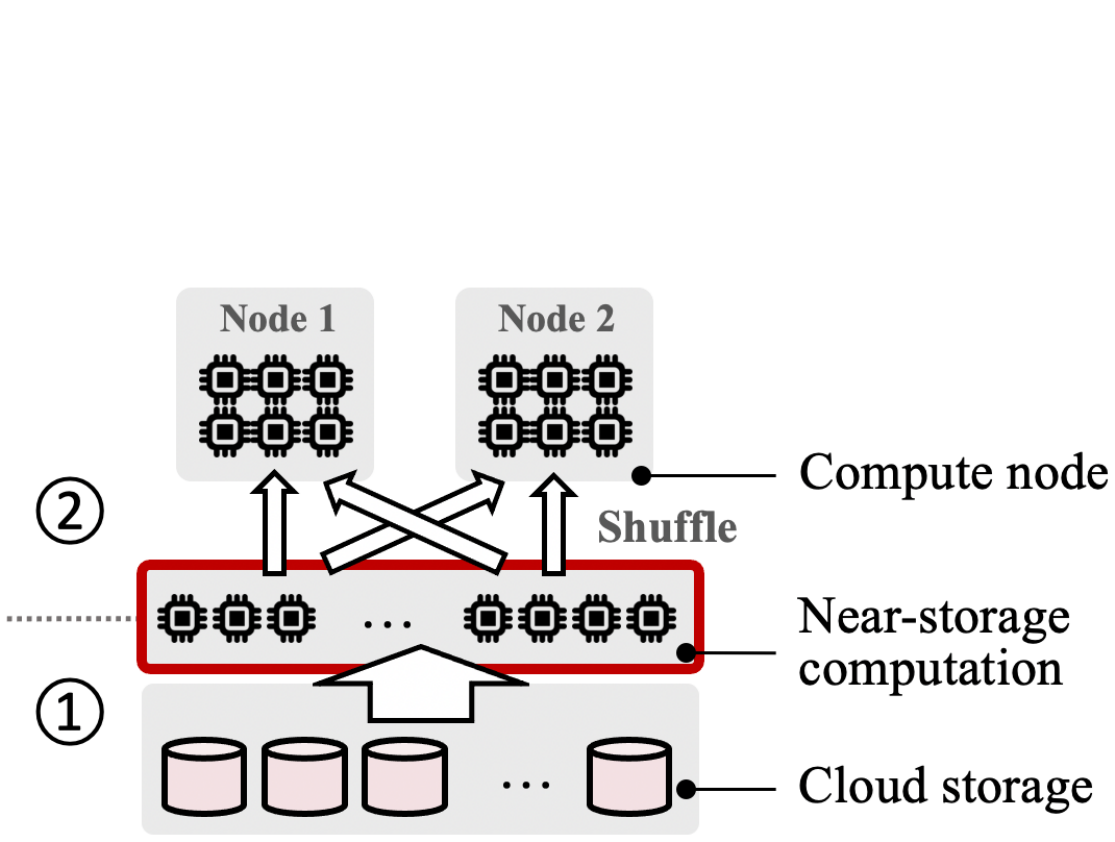}                
    }
    \vspace{-0.2cm}
    \caption{\textbf{Distributed Data Shuffle Pushdown --- \textnormal{Data is directly redistributed to the target compute node from the storage layer.}}}
    \label{fig:shuffle-pd}
\end{figure}

\spara{Distributed Data Shuffle} Traditionally, distributed data shuffle is executed in the computation cluster. This usually happens in a situation when data is loaded from the storage layer and a redistribution step is involved next. For example, when the downstream operator is a hash-join and the two joining relations need to be redistributed based on the join key. Figure~\ref{fig:shuffle-pd}(a) demonstrates the data flow when shuffle is not offloaded to the storage layer. The raw data is initially loaded from the storage devices and processed in the storage layer (Step 1, e.g. selection and projection pushdown). The results are then sent back to the compute nodes as input of the shuffle operation (Step 2). Then the data is partitioned (hash-based, range-based, etc) on the partition key and redistributed to the appropriate compute nodes (Step 3).

Figure~\ref{fig:shuffle-pd}(b) presents the proposed design where we push the shuffle operators into the storage layer. In this case, the processing of the pushdown tasks is initiated on the storage servers without shuffling (Step 1). Before returning the pushdown results to the computation layer, the data is partitioned and directly forwarded to the appropriate target computation nodes (Step 2). Essentially, the new shuffle design merges steps 1 and 2 in the previous design into a single step and execute in the storage layer.

The computation layer needs to send some key parameters to the storage layer for it to conduct the shuffle operator. These include a partition function, the partition key, and the identifiers of the target compute nodes (such as IP addresses and keys) which the shuffled results are returned to. In our implementation, pushdown requests are sent per data partition, and the parameters used in shuffle processing are attached to each pushdown request when sent to the storage layer. For example, assume there are four compute nodes and eight data partitions in the storage layer. Each compute node would have two corresponding data partitions, and thus send two pushdown requests accordingly. The shuffle operation for a data partition is initiated once its upstream operators are finished (e.g. scan, filter).

There are two approaches for a storage server to transfer shuffled data to target compute nodes: (1) actively pushing the data to the compute nodes, or (2) buffering the shuffled results locally and waiting for compute nodes to request. We chose the latter approach since the target compute nodes may not be immediately ready to receive the data when the storage server issues the transmission. However, the storage server has limited memory space and should not indefinitely write shuffled results to the local buffer. The storage server will set an upper bound on the local buffer size; when the local buffer is full, the shuffle operator will throttle until the local buffer is drained by the target computation node.

Pushdown of distributed data shuffle is \textit{local} since it does not incur network traffic across the storage servers --- data is solely transferred from storage to the computation layer. It is also \textit{bounded} because essentially it involves scanning of the input data and assigning records to their corresponding partitions, which consumes CPU and memory resource linearly.

\sspara{Interact with Cached Data} It is non-trivial to exploit the data in the cache while performing shuffle pushdown, since the shuffle operation changes the data distribution, which means the cached data may not be directly applicable to downstream operators.

The most straight forward method is to ignore the cached data when pushing shuffle to the storage layer --- the entire table is redistributed to the compute nodes from the storage. A better solution is to perform the same shuffle function to the cached columns within the compute cluster, and only brings back shuffle results on uncached columns from the storage. The advantages are two-fold. First, in a $n$-node cluster, a portion of roughly $\frac{1}{n}$ network traffic of redistributed data can be saved. To explain, assume the raw data is initially uniformly distribute into the $n$ nodes and the shuffle function evenly partitions the data across the cluster. Then in each node, around $\frac{1}{n}$ of the data will be redistributed to itself, and the rest is forwarded to other nodes. Second, the network bandwidth within the compute cluster is usually higher than the bandwidth between the computation and storage layer, and reading cached data is more efficient than loading data from the storage devices.

However, it may not be applicable to perform the shuffle function over the cached data, which depends on the existence of the shuffle columns in the cache. To tackle this, we can apply the similar idea of selection bitmap pushdown. When processing the partitioning function, the storage layer can generate a \textit{position vector}, which represents the compute node that each row should be redistributed to. For a $n$-node compute cluster, each position value requires $log_2n$ bits, making the position vector a lightweight data structure.

\sspara{Interact with the Upstream Operator} The rationale above can be generalized when the shuffle operation is not the direct downstream operator of data scan --- it may be performed on intermediate results of computation like filtering, aggregation, and join. The input data produced by the upstream operator can be divided into local portion --- produced within the compute layer, and remote portion --- produced within the storage layer (i.e. pushdown), which can be regarded as cached data and uncached data respectively.

\section{Implementation} \label{sec:impl}

In this section, we discuss additional implementation details that have not been discussed in previous sections.

\subsection{DB Engine and Storage Layer}

\name is an open-source cloud-native OLAP DBMS. The latest \name includes both a computation layer and a storage layer. The computation layer is derived from~\cite{fpdb} with several key enhancements, including integration with Apache Calcite~\cite{calcite} as a query optimizer, support for distributed query processing by spawning actors~\cite{caf} remotely across computation nodes at runtime, and enrichment of its supported operators (outer join, top-K, etc.) and expressions (case, string functions, date functions, etc.).

The latest \name also includes a storage layer. To enable customized pushdown features, we develop an open-source storage layer with pushdown capabilities. Data objects are stored on file systems located on locally attached SSDs, which could be accessed by compute nodes through gRPC~\cite{grpc} calls.

\subsection{Pushdown Execution}

At the planning time, \name performs a tree traversal over the query plan from the optimizer. From the leaf nodes (i.e. scan), the pushdown portion expands until reaching an operator (e.g. join) that cannot be executed at storage.

Arrow Flight~\cite{arrow_flight} is leveraged for transferring uncompressed Arrow data from the storage to the compute nodes. Arrow Flight enables wire-speed, zero-copy, and serialization-free data transfer by sending data in Arrow IPC format~\cite{arrow_ipc}, which can be processed directly by \name's executors. Pushdown requests are encapsulated in Arrow Flight Tickets and sent to the storage nodes to process. Each pushdown request contains a serialized query plan instead of a plain SQL dialect~\cite{s3select, azure-accel}, to avoid redundant query parsing and planning at storage. Moreover, we observe inefficiency when transferring large data between actors residing in different nodes, and therefore also utilize Arrow Flight for data transfer within the compute cluster.

For pushdown operators with bitmaps (selection bitmap, bloom filter), \name wraps bitmaps into single-column Arrow tables for low serialization overhead. In adaptive pushdown, when the storage decides to push back a request, a special Flight error will be returned to the computation layer, where the corresponding compute node will issue a another RPC call to retrieve the compressed raw data.

\section{Evaluation} \label{sec:eval}

In this section, we evaluate the performance of adaptive pushdown and new pushdown operators using standard OLAP benchmarks.

\subsection{Experimental Setup} \label{sec:eval-setup}

\spara{Hardware Configuration} We conduct all the experiments on AWS EC2 virtual machines. We use r5.4xlarge instances (which costs \$1.008 per hour in US-East-2 pricing) with 16~vCPU, 128~GB memory, and up to 10~Gbps network bandwidth in the computation layer. The storage layer is set up on r5d.4xlarge instances (which costs \$1.152 per hour in US-East-2 pricing), with 16~vCPU, 128~GB memory, up to 10~Gbps network bandwidth, and two 300~GB local NVMe SSDs. All servers run the Ubuntu 20.04 operating system. For experiments with distributed data shuffle pushdown, we configure both the compute layer and the storage layer to 4-node clusters. Otherwise, we configure both layers to a single instance.

\spara{Benchmark} We use the widely adopted data analytics benchmark, \textit{TPC-H}~\cite{tpch}. TPC-H contains 22 queries in total. We use a scale factor of 100 (100~GB data set when uncompressed) for experiments with distributed data shuffle pushdown, and a scale factor of 50 (50~GB data set when uncompressed) otherwise. Each table is sharded into partitions of roughly 150~MB when using CSV format. Table partitions are converted to Parquet~\cite{parquet} format and evenly distributed to local disks of the storage instances. All the experiments are conducted using the Parquet data.

\spara{Measurement} We measure the execution time and other relevant metrics (e.g., network traffic, number of admitted requests, etc.) for each experiment. For each query, we run three times and record the average execution time. Since selection bitmap pushdown needs to interact with the local cache of \name, we conduct separate experiments to evaluate its performance. Specifically, each experiment consists of a warm-up query that populates the data into the cache, and an execution query where the performance is measured.
\subsection{Evaluating Adaptive Pushdown} \label{sec:eval-adapt}

\begin{figure*}[ht]
    \subfloat{
        \includegraphics[width=\linewidth]{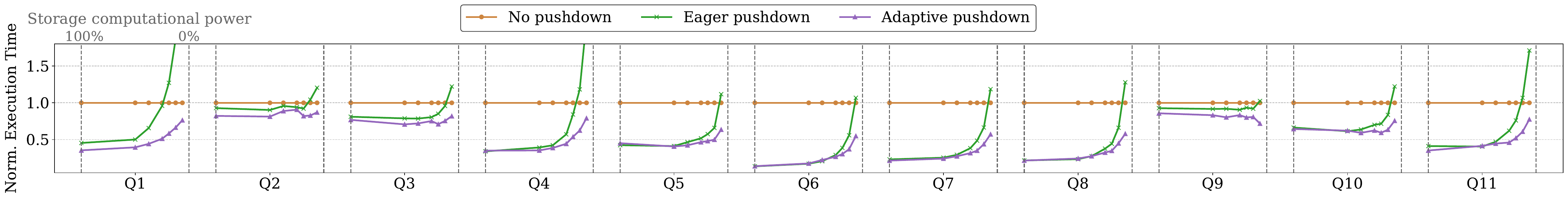}                
    }
    \\
    \vspace{-0.35cm}
    \subfloat{
        \includegraphics[width=\linewidth]{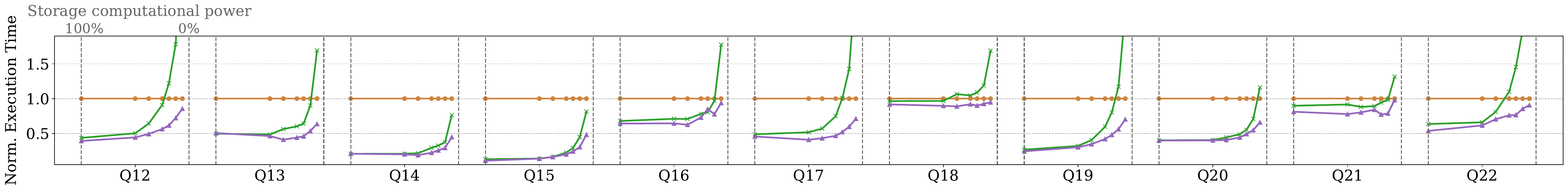}
    }
    \vspace{-0.35cm}
    \caption{Performance Evaluation of Adaptive Pushdown on TPC-H \textnormal{(normalized to \textit{No pushdown})}.
    % \aia{Consider shrinking the range of the y-axis to emphasize the difference between eager and adaptive pushdown. Currently there is a lot of vertical empty space in the figures.}
    }
    \label{fig:exp-adapt-main}
\end{figure*}

In this section, we evaluate the performance of adaptive pushdown in different storage-layer computational resource conditions. This can be affected by both the CPU power of the storage server, and its computational resource usage. We emulate the storage-layer computational resource status by varying the number of available CPU cores for pushdown tasks (with a storage computational power of 1 meaning that all CPU cores are available). Since \name is built on top of the actor framework~\cite{caf}, we achieve this by configuring the maximal number of threads that can be exploited by the actor system scheduler. At each time we measure the performance of a single query.
% \msc{My understanding is that you run one query at a time and vary the amount of resources available to simulate resource contention. Is that correct? The language is not really clear.}

\spara{Overall Performance} Figure~\ref{fig:exp-adapt-main} compares the execution time of \np, \ep, and \ap under different computational resource utilization status in the storage. Results are normalized to \np. When the computational resource at storage is abundant for pushdown execution (i.e., storage computational power is higher than 0.5), \ep outperforms \np and is only slightly affected by the computational power at storage. As the storage computational power decreases, pushdown execution gets throttled and gradually becomes the major bottleneck, making \ep underperform \np when the storage-layer computational resource is scarce.

The performance of \ap is consistently better than both baselines. Specifically, when the storage computational power is high, it performs similarly to \ep, and when the storage server is tiny or under heavy burden, its performance degrades less than \ep and can still slightly outperform \np. In situations where the storage computational power falls between these two extremes, adaptive pushdown achieves the best of both worlds. When the performance of \np and \ep breaks even, \ap outperforms both baselines by 1.5$\times$ on average, and queries including Q1, Q6, Q8, Q17, and Q19 achieve a speedup of 1.9$\times$.

With pushdown enabled, the sensitivity on the storage-layer CPU utilization varies among different queries. Most queries (15 of all) expose a high sensitivity when executed with pushdown. For example, the performance of \ep of Q1, Q12, Q19, and Q22 is greatly impact by the storage-layer computational power, and starts to degrade even when the storage-layer CPU resource is not scarce. In these queries, the performance improvement of \ap is prominent since the pushable portion of the query plan dominates the end-to-end execution time, and \ap mitigates the issue of resource contention at storage when the pushable subquery plan is processed. For the remaining queries, the performance of \ep is not very sensitive to the storage-layer computational power (e.g., Q2, Q3, and Q18), where the execution time is dominated by non-pushable operators. In these queries, \ap only shows its superiority when the available computational resource at storage is extremely low. For example, \ap in Q2 outperforms both baselines by 1.2$\times$ when the storage-layer computational power is 25\%.

\spara{Case Study} To get a deeper understanding of the performance benefits, we conduct detailed analysis on two representative queries. We pick Q14 as the representative query that benefits significantly from computation pushdown, and Q12 as the representative query that benefits less. 

\begin{figure}[h]
\begin{minipage}{0.6\linewidth}
    \centering
    \includegraphics[width=\linewidth]{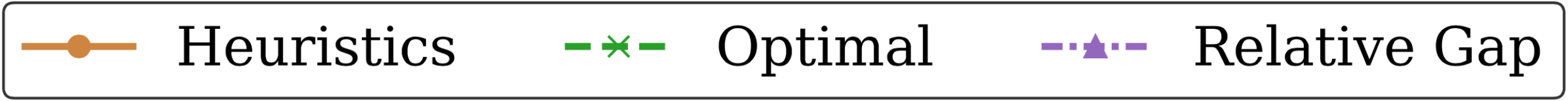}
\end{minipage}\\
\vspace{-0.12in}
\begin{minipage}{\linewidth}
    \subfloat[Q12 \textnormal{(Total = 190)}]{
        \includegraphics[width=0.49\linewidth]{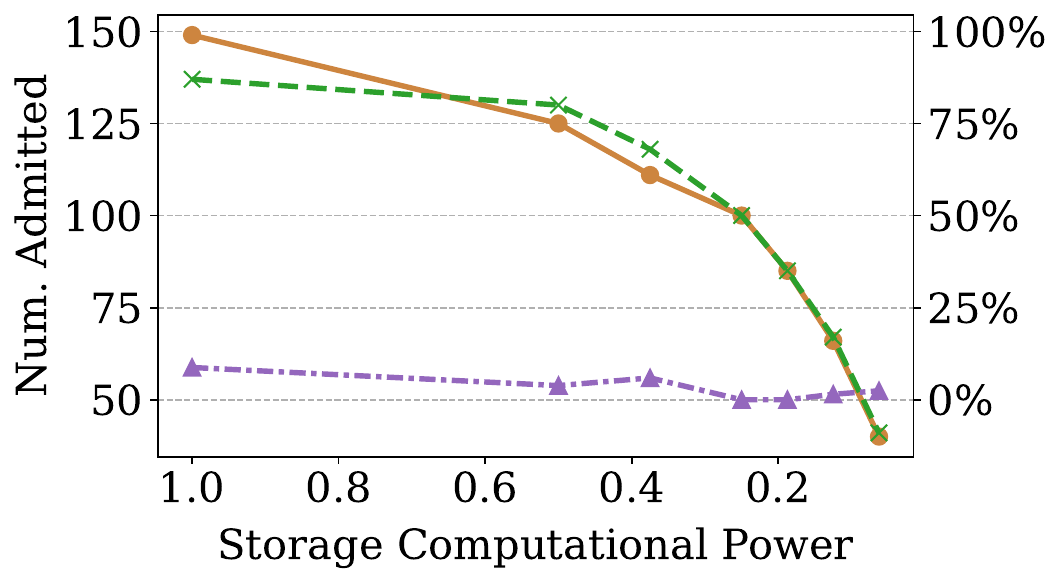}
    }
    \hspace{-0.2cm}
    \subfloat[Q14 \textnormal{(Total = 156)}]{
        \includegraphics[width=0.49\linewidth]{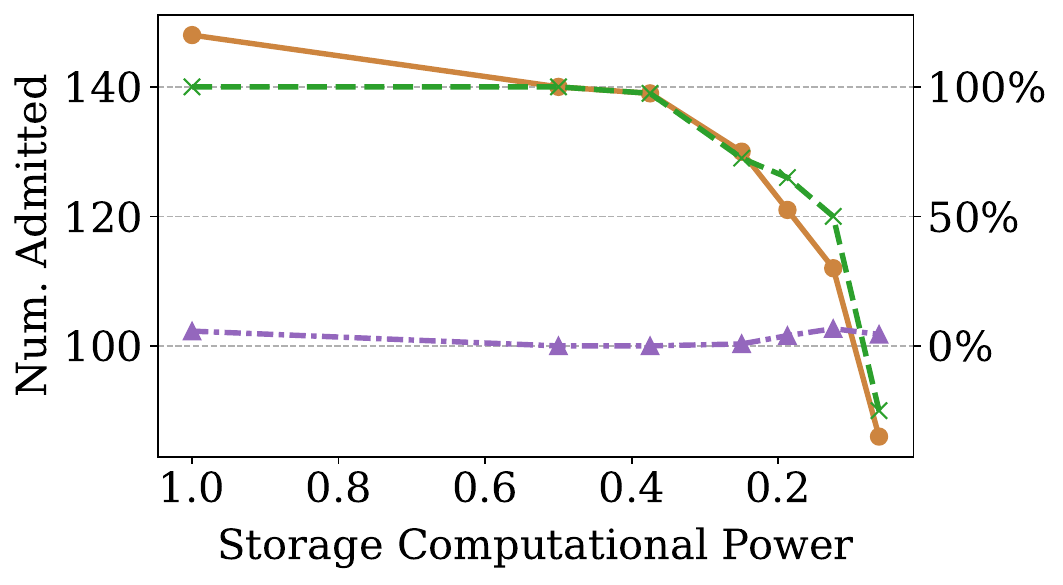}
    }
\end{minipage}
\vspace{-0.3cm}
\caption{Comparison between Pushback Heuristics and the Theoretical Optimal Bound.}
\label{fig:exp-adapt-pb}
\end{figure}
\begin{figure}[h]
\begin{minipage}{0.8\linewidth}
    \centering
    \includegraphics[width=\linewidth]{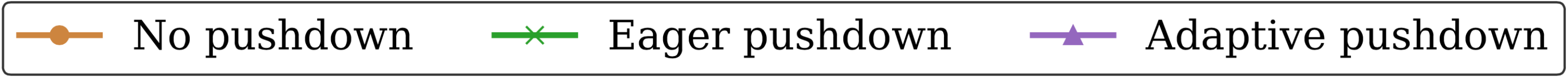}
\end{minipage}\\
\vspace{-0.12in}
\begin{minipage}{\linewidth}
    \subfloat[Q12]{
        \includegraphics[width=0.48\linewidth]{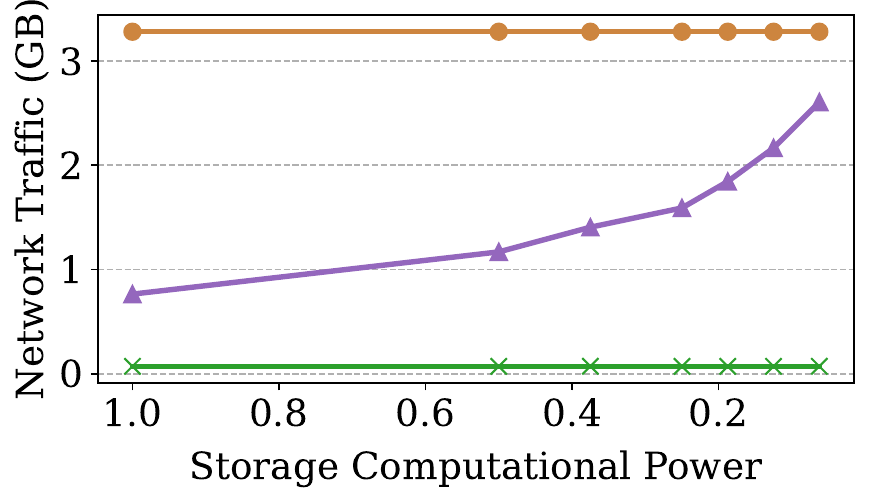}
    }
    \hspace{-0.1cm}
    \subfloat[Q14]{
        \includegraphics[width=0.48\linewidth]{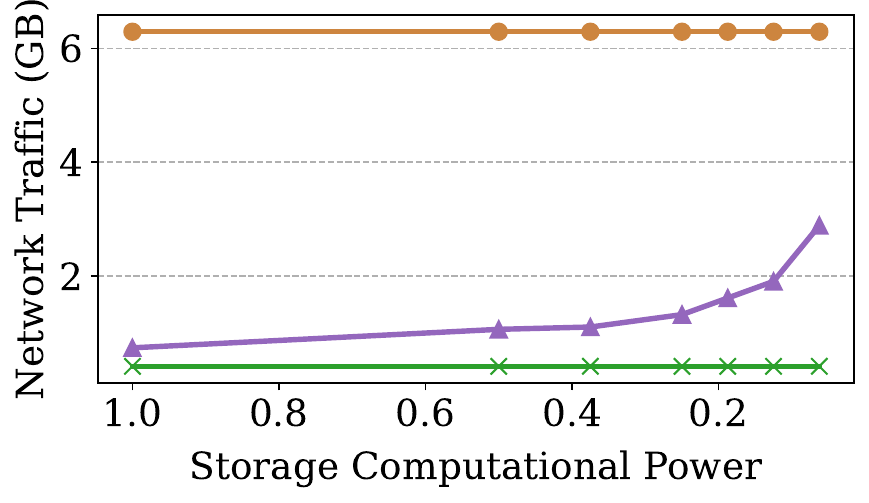}
    }
\end{minipage}
\vspace{-0.3cm}
\caption{Network Traffic Measured on Two Representative Queries \textnormal{(Q12 and Q14)}.}
\label{fig:exp-adapt-net}
\end{figure}
\begin{figure}[h]
\begin{minipage}{0.8\linewidth}
    \centering
    \includegraphics[width=\linewidth]{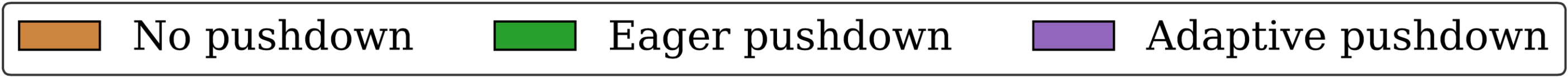}
\end{minipage}\\
\vspace{-0.03cm}
\begin{minipage}{0.93\linewidth}
    \centering
    \includegraphics[width=\linewidth]{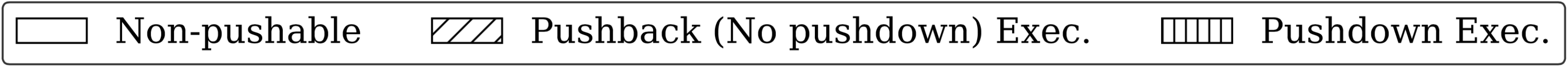}
\end{minipage}\\
\vspace{-0.12in}
\begin{minipage}{\linewidth}
    \subfloat[Q12]{
        \includegraphics[width=0.48\linewidth]{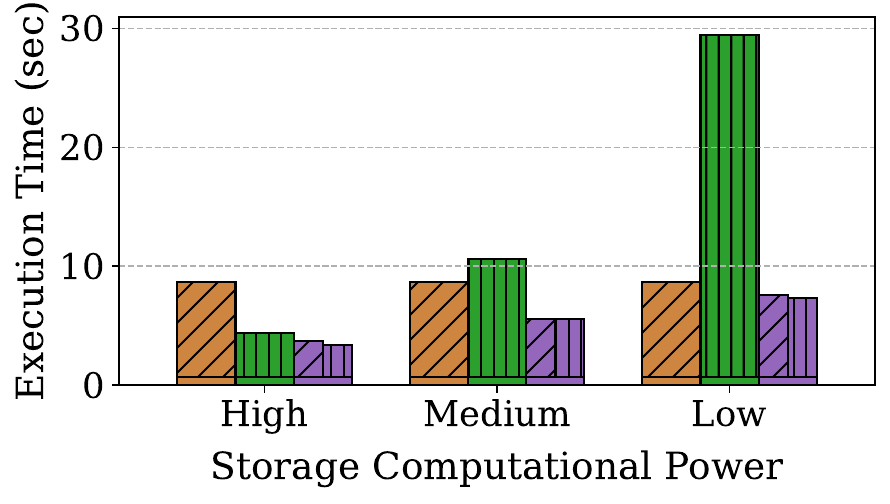}
    }
    \hspace{-0.1cm}
    \subfloat[Q14]{
        \includegraphics[width=0.48\linewidth]{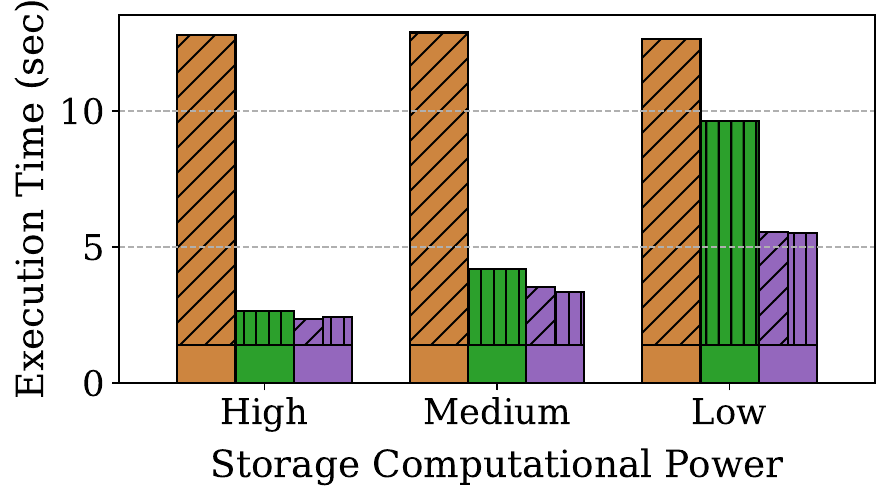}
    }
\end{minipage}
\vspace{-0.3cm}
\caption{Performance Breakdown on Two Representative Queries \textnormal{(Q12 and Q14)}.}
\label{fig:exp-adapt-break}
\end{figure}

We measure the number of admitted pushdown requests at the storage layer in each experiment. Figure~\ref{fig:exp-adapt-pb} shows the results of the heuristics used in the pushback mechanism (Algorithm~\ref{alg:mech}). For both queries, with the the storage computational power decreasing, fewer pushdown requests are admitted to be executed on the storage server, and more requests are pushed back to the compute layer. Compared to Q12, pushbacks in Q14 are less frequent since it achieves a higher maximal pushdown speedup, such that more tasks are executed in the storage.

We further evaluate the gap between the pushback heuristics and the theoretical optimal bound (Section~\ref{sec:adapt-theo}), by comparing the number of actual admitted pushdown requests in the storage with the theoretical result obtained from Equation~\ref{eq:theo-opt-5}. Overall we observe a very small relative gap between the heuristics and optimal (1\% on Q12 and 2\% on Q14), and in some cases pushback heuristics achieve the optimal bound (e.g., for Q12 when the storage computational power is less than 25\%). This demonstrates that the pushback mechanism is able to find a proper division of the computation tasks between pushdown and non-pushdown.

Figure~\ref{fig:exp-adapt-net} compares the incurred network traffic between the storage and compute layers among different pushdown strategies. The network traffic of \np and \ep both remain consistent, and \ep reduces network traffic up to an order of magnitude. The network usage of \ap is sensitive to the storage-layer computational power, since it adaptively adjusts the ratio between assigned pushdown and pushback tasks, such that both CPU and network resources at the storage server can be fully utilized.

% \yxy{Is "overhead" in the y-axis the same as runtime? If so, maybe use runtime to avoid confusion.}
Figure~\ref{fig:exp-adapt-break} shows the performance breakdown on the two representative queries. We present three cases where the storage-layer computational power is high, medium, and low, respectively. The execution time of the non-pushable portion of the query plan remains stable in all cases. Compared to fetching raw data from the storage in \np, the overhead of pushback executions in \ap is consistently smaller. Similarly, the execution time of pushdown executions in \ap is always lower than it in \ep. Additionally, in \ap pushdown and pushback executions happen in parallel, and the performance is determined by the slower execution path. In both queries we observe a similar execution time between pushdown and pushback executions, denoting our algorithm is able to find a proper division between pushdown and pushback tasks and balance the usage of CPU and network resources.

\spara{Awareness of Pushdown Amenability} Next we evaluate \ap when the pushdown requests have different pushdown amenabilities. Within a single query, pushdown amenability typically appears to be similar, since TPC-H dataset embraces a uniform distribution among different data partitions. Therefore in this experiment, we execute multiple queries simultaneously. Specifically, we pick a combination of two queries, Q12 and Q14, where one query (Q14) benefits significantly from pushdown and the other query (Q12) is less pushdown-amenable.

\begin{figure}[h]
\begin{minipage}{0.7\linewidth}
    \centering
    \includegraphics[width=\linewidth]{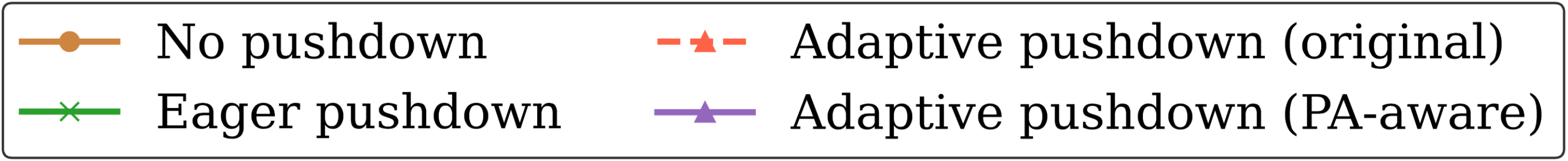}
\end{minipage}\\
\vspace{-0.12in}
\begin{minipage}{\linewidth}
    \subfloat[Q12]{
        \includegraphics[width=0.48\linewidth]{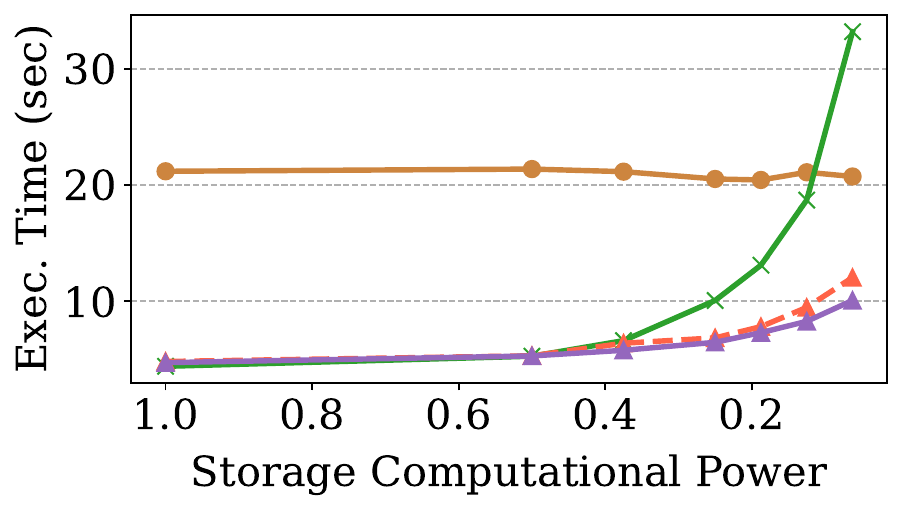}
    }
    \hspace{-0.1cm}
    \subfloat[Q14]{
        \includegraphics[width=0.48\linewidth]{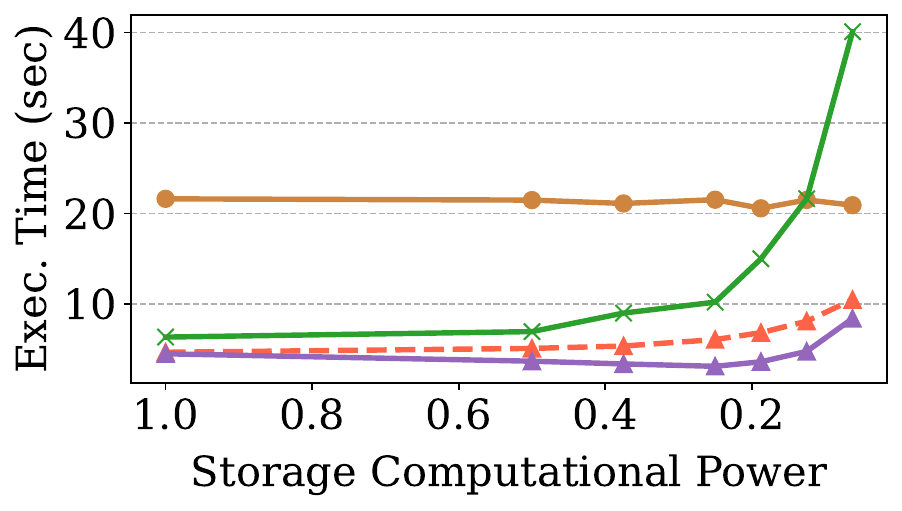}
    }
\end{minipage}
\vspace{-0.3cm}
\caption{Evaluation of Awareness of Pushdown-Amenability in Concurrent Executions.}
\label{fig:exp-adapt-pa}
\end{figure}
\begin{figure}[h]
\begin{minipage}{0.8\linewidth}
    \centering
    \includegraphics[width=\linewidth]{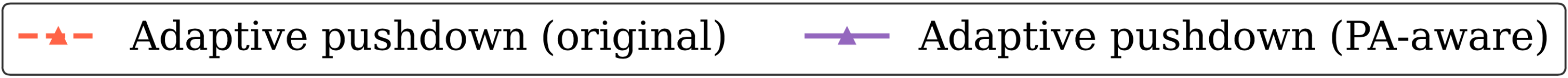}
\end{minipage}\\
\vspace{-0.12in}
\begin{minipage}{\linewidth}
    \subfloat[Q12 \textnormal{(Total = 190)}]{
        \includegraphics[width=0.48\linewidth]{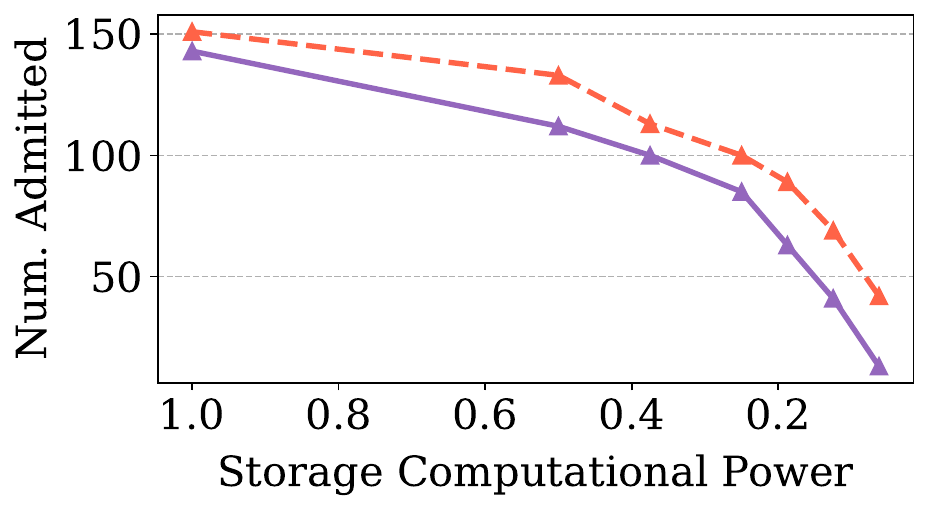}
    }
    \hspace{-0.1cm}
    \subfloat[Q14 \textnormal{(Total = 156)}]{
        \includegraphics[width=0.48\linewidth]{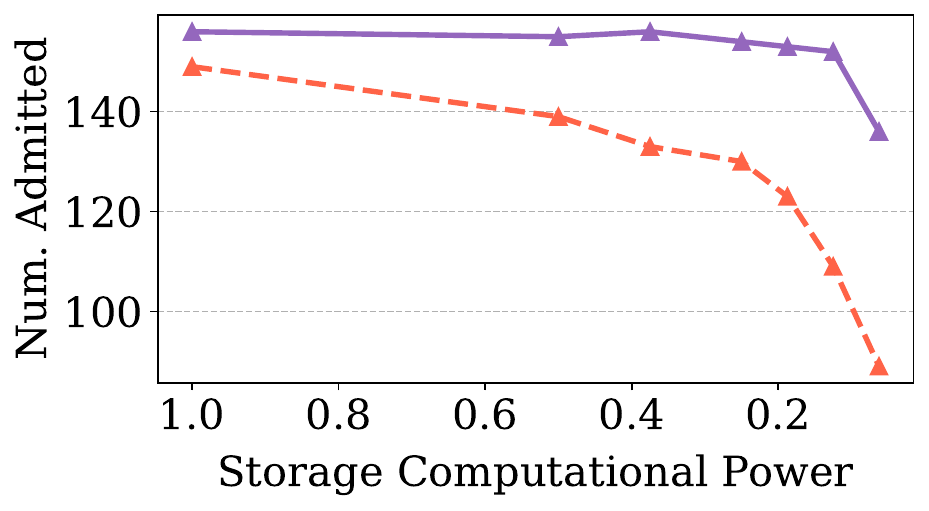}
    }
\end{minipage}
\vspace{-0.3cm}
\caption{Number of Admitted Pushdown Requests at storage in Concurrent Executions.}
\label{fig:exp-adapt-pa-pb}
\end{figure}
\begin{figure}[h]
\begin{minipage}{0.7\linewidth}
    \centering
    \includegraphics[width=\linewidth]{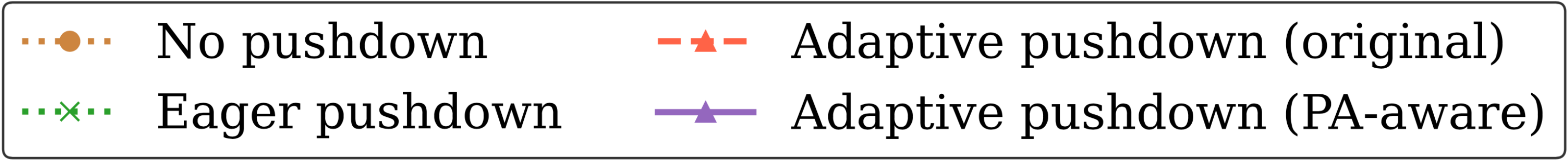}
\end{minipage}\\
\vspace{-0.12in}
\begin{minipage}{\linewidth}
    \subfloat[CPU]{
        \includegraphics[width=0.477\linewidth]{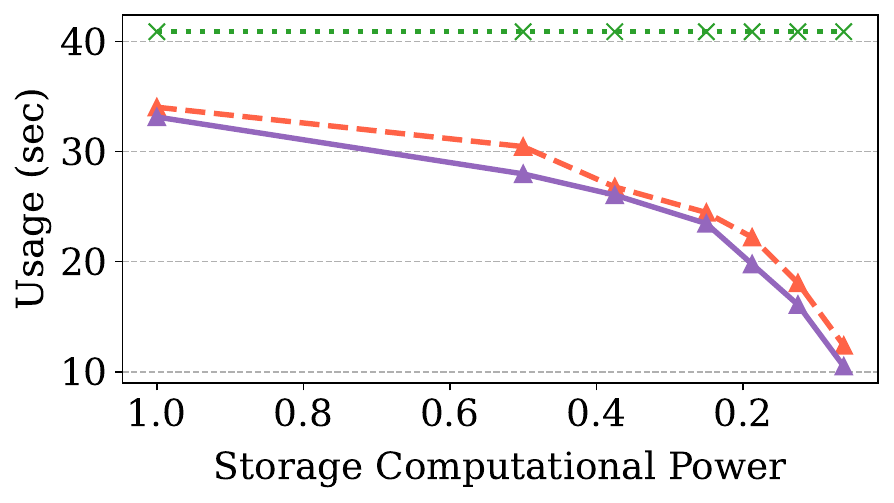}
    }
    \hspace{-0.1cm}
    \subfloat[Network]{
        \includegraphics[width=0.483\linewidth]{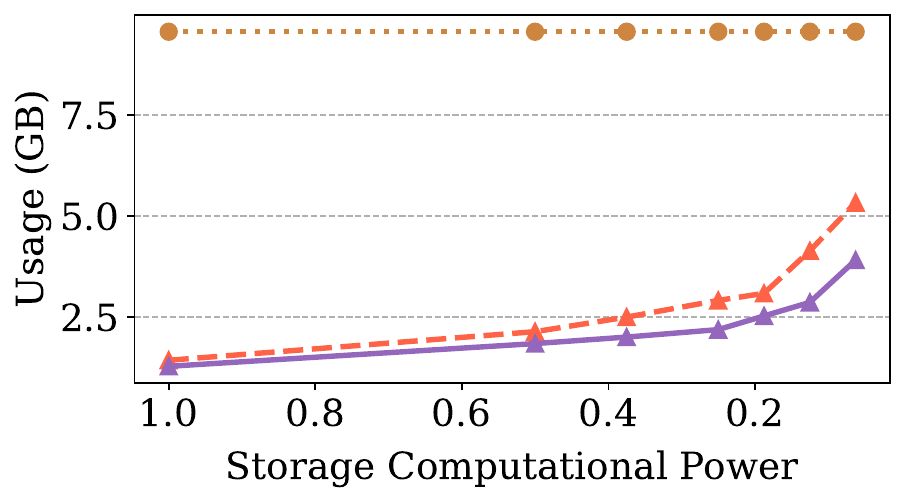}
    }
\end{minipage}
\vspace{-0.3cm}
\caption{Resource Usage in Concurrent Executions \textnormal{(CPU usage is measured by the total CPU time that is normalized to the time of 1 vCPU)}.}
\label{fig:exp-adapt-pa-usage}
\end{figure}

Figure~\ref{fig:exp-adapt-pa} compares the original \ap and \ap that is aware of pushdown amenability. \np and \ep baselines are also added. Both original \ap and Pushdown-Amenability-aware (\textit{PA-aware} in short) \ap outperform the two baselines. Compared to the original \ap, PA-aware \ap further improves the performance of both concurrent queries, where Q12 is accelerated by up to 1.2$\times$ and Q14 is improved by up to 1.9$\times$. An interesting observation on Q14 is that the performance is sometimes even improved with lower storage computation power (e.g., 0.3). This is because the performance gap between the two concurrent queries are increased, such that the contention on the non-pushable portion in the compute layer is mitigated --- the slower query (Q12) has not entered the non-pushable portion when the faster query (Q14) has completed.
% \yxy{Figure 11 suggests that lower storage computation power can lead to lower runtime for PA-aware. Maybe double check this.}

To understand the achieved performance improvement, we trace the number of admitted pushdown requests for both queries respectively, which is shown in Figure~\ref{fig:exp-adapt-pa-pb}. Overall, we observe a decrease for the number of admitted pushdown requests on Q12 but an increase on Q14. This is because the requests of Q14 have a potentially larger pushdown benefit, and they are prioritized to be executed at the storage layer. Correspondingly, more requests of Q12 are pushed back to the compute layer. It is interesting to note that the performance of Q12 does not degrade but is even slightly improved, where the reasons are two-fold. First, the difference of the execution time between the pushback path and pushdown path on Q12 is not significant, so a small number of more pushback execution do not hurt the performance. Second, since the requests of Q14 are executed more efficiently, the time spent in the wait queue for the requests of Q12 decreases.

We further investigate the resource usage of CPU and network, which is demonstrated in Figure~\ref{fig:exp-adapt-pa-usage}. \np and \ep baselines are added for reference. PA-aware \ap reduces the CPU usage by up to 15\%, and network usage by up to 31\%, compared to the original mechanism. The reduction is more significant when the storage-layer computational power is lower, since more requests are pushed back to the compute layer, and PA-aware \ep is able to capture the most proper requests that should be pushed back.
\subsection{Evaluating Pushdown Operators}

In this section, we evaluate the performance of new pushdown operators, namely, selection bitmap and distributed data shuffle. We implement pushdown operators in existing systems (selection, projection, aggregation, bloom filter, etc.) as the baseline.

\subsubsection{Evaluating Selection Bitmap Pushdown}
\hfill\\
We conduct experiments on several representative benchmark queries: Q3, Q4, Q12, Q14, and Q19 (other queries observe similar results). In each experiment, we vary the selectivity of the filter predicate associated with the fact table \textit{Lineitem}.

\begin{figure}[h]
\begin{minipage}{0.6\linewidth}
    \centering
    \includegraphics[width=\linewidth]{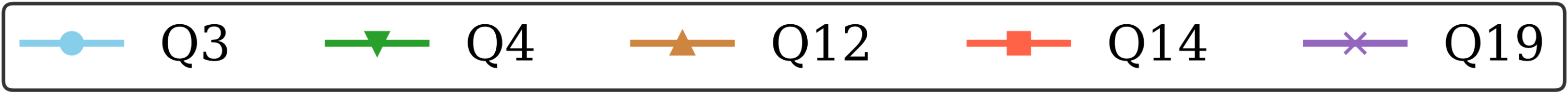}
\end{minipage}\\
\vspace{-0.15in}
\begin{minipage}{\linewidth}
    \subfloat[Execution Time]{
        % \raisebox{0.07cm}{
            \includegraphics[width=0.465\linewidth]{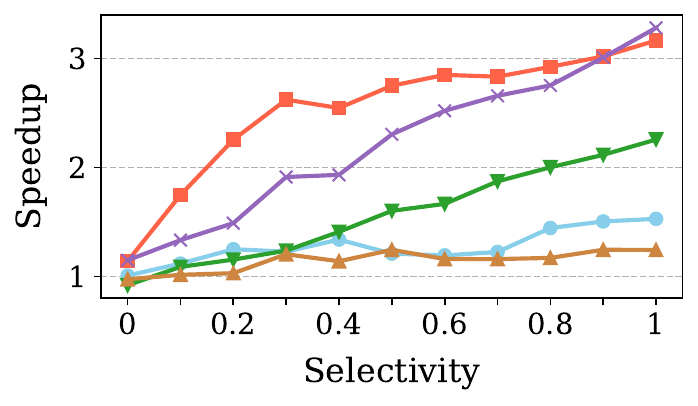}                
        % }
    } 
    \subfloat[Network Traffic]{
        \includegraphics[width=0.495\linewidth]{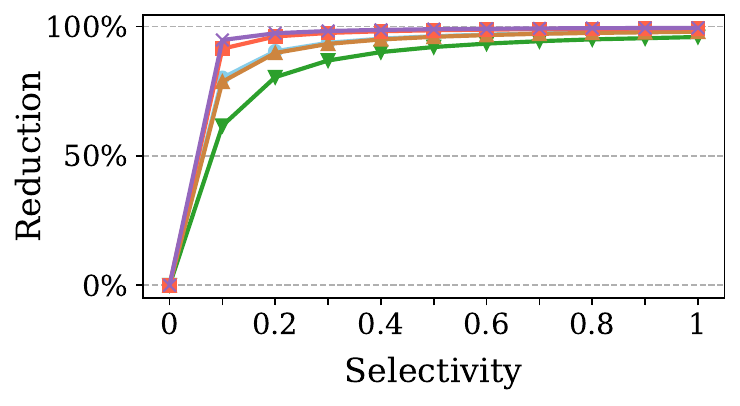}
    }
\end{minipage}
\vspace{-0.3cm}
\caption{Performance Evaluation of Selection Bitmap Pushdown \textnormal{(Results are normalized to \textit{Pushdown (baseline)})} --- \normalfont{The selection bitmap is constructed at the storage layer.}}
\label{fig:exp-bitmap-storage}
\end{figure}

\spara{Selection Bitmap from the Storage Layer} We first simulate the scenario where the selection bitmap can only be generated in the storage layer. We achieve this by caching only the output columns of the filter operator for the fact table. Predicate columns are not cached. The results are depicted in Figure~\ref{fig:exp-bitmap-storage}.

\vspace{0.1cm}

\vspace{-0.6cm}
\begin{figure}[h]
\hspace{-0.3cm}
\begin{minipage}{\linewidth}
    \subfloat[Execution Time]{
        \includegraphics[width=0.48\linewidth]{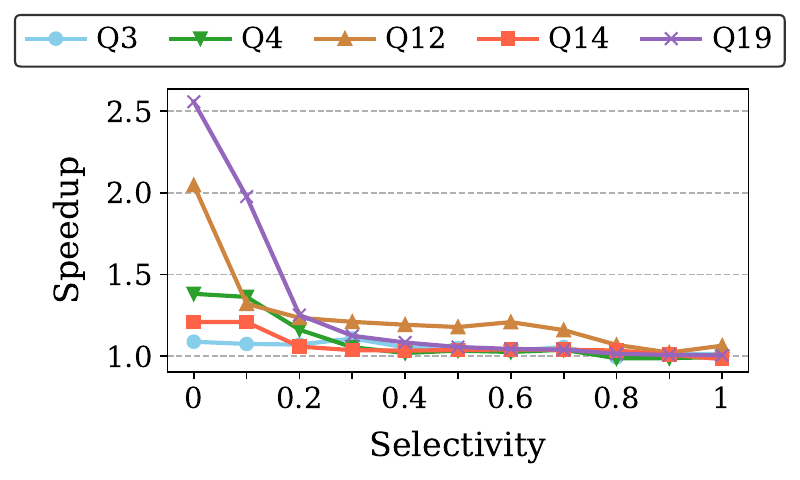} 
    }
    \subfloat[Disk Metrics]{
        \raisebox{0.005cm}{
        \includegraphics[width=0.48\linewidth]{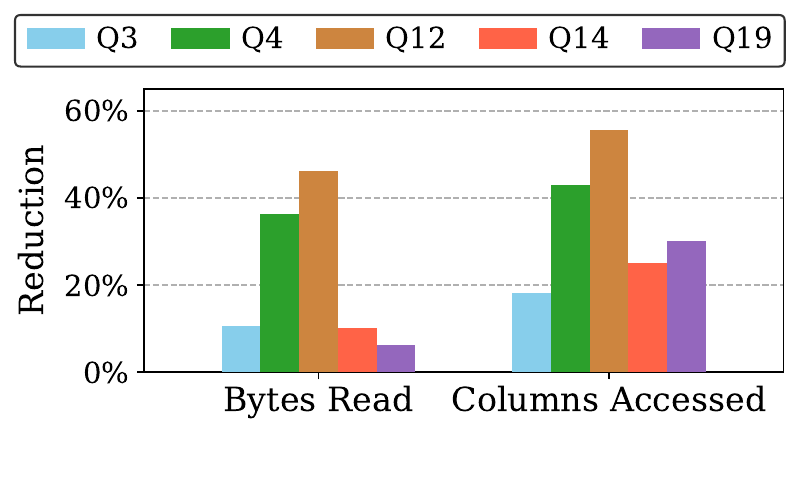}
        }
    }
\end{minipage}
\vspace{-0.3cm}
\caption{Performance Evaluation of Selection Bitmap Pushdown \textnormal{(Results are normalized to \textit{Pushdown (baseline)}) --- The selection bitmap is constructed at the computation layer.}}
\label{fig:exp-bitmap-compute}
\end{figure}
% \begin{figure*}[ht]
%     \subfloat[Execution Time \textnormal{(normalized to \textit{No pushdown})}]{
%         \includegraphics[width=\linewidth]{figures/eval/runtime_shuffle.pdf}                
%     }
%     \\
%     \vspace{-0.3cm}
%     \subfloat[Network Traffic \textnormal{(normalized to \textit{No pushdown})}]{
%         \includegraphics[width=\linewidth]{figures/eval/network_shuffle.pdf}
%     }
%     \vspace{-0.3cm}
%     \caption{Performance Evaluation of Distributed Data Shuffle Pushdown on TPC-H.}
%     \label{fig:exp-shuffle}
% \end{figure*}

\begin{figure*}[ht]
    \includegraphics[width=\linewidth]{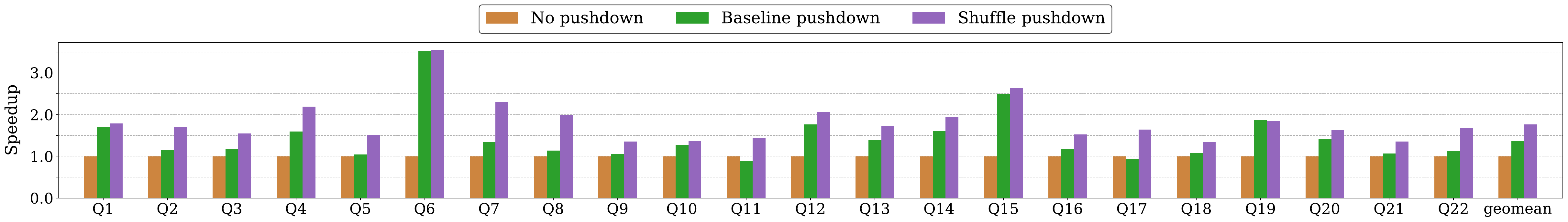}
    \vspace{-0.7cm}
    \caption{Performance Evaluation of Distributed Data Shuffle Pushdown on TPC-H \textnormal{(normalized to \textit{No pushdown})}.}
    \label{fig:exp-shuffle}
\end{figure*}

As Figure~\ref{fig:exp-bitmap-storage}(a) illustrates, all queries show an improvement in performance compared to the baseline. Selection bitmap pushdown is most effective in Q14 and Q19. When the filter predicate is non-selective (e.g., 0.9), these two queries perform over 3.0$\times$ better than baseline pushdown since transferring selection bitmaps instead of data columns reduces network traffic significantly (over 90\% of data transfer is saved, as shown in Figure~\ref{fig:exp-bitmap-storage}(b)). Q4 also observes a speedup of up to 2.3$\times$.

When the filter predicate is highly selective, the speedup is less obvious. This is because baseline pushdown returns less data to the computation layer, such that the reduction of data transfer is less significant. Despite this, query execution still gets accelerated. For instance, when the selectivity is 0.1, the speedups of Q14 and Q19 are 1.8$\times$ and 1.3$\times$ respectively.

The performance gain on Q3 and Q4 is less substantial compared to the other queries. These queries contain more complex operators downstream of pushdown (e.g., more intricate joins and aggregations), leading to a diluted performance benefit.

\spara{Selection Bitmap from the Computation Layer} We next emulate the case where the compute-layer selection bitmap can be used to accelerate pushdown execution in the storage layer. In this experiment, only the predicate columns are cached in the compute nodes. Results are displayed in Figure~\ref{fig:exp-bitmap-compute}.

As demonstrated in Figure~\ref{fig:exp-bitmap-compute}(a), all queries benefit from selection bitmap pushdown when the filter predicate is selective. For example, when the selectivity approaches 0, pushdown of selection bitmaps outperforms the baseline by 2.0$\times$ and 2.6$\times$ on Q12 and Q19, respectively. When the filter predicate becomes less selective, the performance gain decreases, since more data is transferred to the compute nodes, which dominates the query execution time.

We also analyze disk metrics by measuring the number of bytes read and the number of columns accessed from the disks, which are illustrated in Figure~\ref{fig:exp-bitmap-compute}(b). The amount of data scanning is reduced by 36\% and 46\% on Q4 and Q14 respectively, and by approximately 10\% on the rest queries. Additionally, the number of columns accessed of the Parquet data decreases between 18\% and 56\%. The reduction of data scanning is less substantial compared to column access reduction, because the columns that can be skipped via selection bitmap pushdown are typically highly compressed, such as \textit{l\_shipmode}, which only has 7 unique values, and \textit{l\_quantity}, of which the value is within a small range between 1 and 50. Conversely, columns that must be transferred are often join keys or have a decimal type, which usually have a low compression ratio.

\subsubsection{Evaluating Distributed Data Shuffle Pushdown}
\hfill\\
Next we evaluate the performance of distributed data shuffle pushdown over TPC-H, which is shown in Figure~\ref{fig:exp-shuffle}. All execution times are normalized to the \np. Across all queries, shuffle pushdown results in an average of 1.3$\times$ performance improvement over baseline pushdown, and 1.8$\times$ over no pushdown.

Among all 22 queries, we observe the performance improvement on 20 of them, with 15 queries are accelerated by over 1.2$\times$, 10 queries accelerated by over 1.3$\times$, and 6 queries accelerated by over 1.5$\times$. Q7, Q8, and Q17 benefit from shuffle pushdown most significantly, which are improved by more than 1.7$\times$. In these queries, the filter predicates associated with the base tables are not selective, where a major part of the table data still needs to be fetched. Shuffle pushdown is able to eliminate the redistribution of the scanned base table data, which occupies a large portion of the overall execution time. As a result, we observe more than half of the data exchange across the compute nodes is saved compared to baseline pushdown.

Conversely, several queries do not benefit a lot from shuffle pushdown (e.g., Q6, Q15, and Q19). These queries typically have selective filter predicates on base tables, such that the amount of data transferred from the storage layer is not significant, and the overhead of data exchange across the compute nodes is negligible.

We further investigate the incurred network traffic in different pushdown configurations. On average shuffle pushdown reduces the consumed network resource by 38\%. Specifically, shuffle pushdown reduces the data exchange across the compute layer by 84\% on overage, while the network traffic between the compute and storage layers is unaffected. Out of all 22 queries, the incurred data exchange across the compute layer is reduced by over 50\% on 20 queries, by over 90\% on 16 queries, and by over 99\% on 7 queries --- only a few small intermediate join results are redistributed.

\section{Related Work} \label{sec:related}

The concept of computation pushdown has been widely adopted by modern cloud-native databases in a storage-disaggregation architecture, typically in cloud OLAP databases. Examples of these systems include AWS Redshift Spectrum~\cite{spectrum}, S3 Select~\cite{s3select}, and Azure Data Lake Storage Query Acceleration~\cite{azure-accel}. Systems such as Presto~\cite{presto}, PushdownDB~\cite{pushdowndb} and FlexPushdownDB~\cite{fpdb} support computation offloading via S3 Select. PolarDB-X~\cite{polardbx} incorporates more pushdown operators like sorting and co-located equi-joins. These systems use software techniques to implement pushdown functionalities. AWS Advanced Query Accelerator (AQUA)~\cite{aqua} uses special hardware accelerators (AWS Nitro chips~\cite{nitro}) to improve the speed and reduce energy consumption of pushdown functions. OLTP databases in the cloud also embrace pushdown computation. For example, AWS Aurora~\cite{aurora1, aurora2} is deployed on a custom-designed cloud storage layer where functionalities including log replay and garbage collection are offloaded to the storage layer.

Beyond cloud-native databases, computation pushdown has also been investigated in various research fields. In the world of database machines, computation is often offloaded to storage through specialized hardware~\cite{netezza, idm, exadata, grace}. For instance, IBM Netezza data warehouse appliances~\cite{netezza} enable pushdown of selection, projection, and compression to disks via FPGA-enabled near-storage processors. The Intelligent Database Machines (IDM)~\cite{idm} relocate most DBMS functionalities to a backend machine close to the disks. 

The technique of computation pushdown is also investigated in Smart Disks/SSD~\cite{riedel2001active, keeton1998case, summarizer, biscuit, do2013query, hrl, ibex, aquoman}. For example, Active Disks~\cite{riedel2001active} and IDISKS~\cite{keeton1998case} explored offloading computation to magnetic storage devices. Meanwhile, architectures like Summarizer~\cite{summarizer} and Biscuit~\cite{biscuit} support selection pushdown to SSD processors. Other Smart SSDs~\cite{do2013query} and near-storage FPGAs~\cite{hrl, ibex} can handle filtering and aggregation tasks in place.

Moreover, computation pushdown has been explored through processing-in-memory (PIM) techniques that utilize DRAM or NVM. Modern 3D-stacked DRAM has a logic layer below the DRAM cell arrays within the same chip, thereby reducing data transfer between memory and CPU~\cite{ghose2018enabling}. Kepe et al.~\cite{kepe2019database} conduct an experimental study that focuses on selection processing in PIM.

\section{Conclusion} \label{sec:conclu}

This paper presents \ap, a technique that leverages a pushback mechanism to determine whether a pushdown operator should be executed in the storage adaptively based on the storage-layer resource utilization. We additionally conduct a systematical analysis of existing pushdown operators and propose two new operators that can benefit from offloaded to storage. Evaluation shows that \ap and the new pushdown operators lead to up to 1.9$\times$ and 3$\times$ speedup on TPC-H benchmark respectively.

\small
\bibliographystyle{ACM-Reference-Format}
\bibliography{ref}


\begin{thebibliography}{47}

%%% the Web address.

\ifx \showCODEN    \undefined \def \showCODEN     #1{\unskip}     \fi
\ifx \showDOI      \undefined \def \showDOI       #1{#1}\fi
\ifx \showISBNx    \undefined \def \showISBNx     #1{\unskip}     \fi
\ifx \showISBNxiii \undefined \def \showISBNxiii  #1{\unskip}     \fi
\ifx \showISSN     \undefined \def \showISSN      #1{\unskip}     \fi
\ifx \showLCCN     \undefined \def \showLCCN      #1{\unskip}     \fi
\ifx \shownote     \undefined \def \shownote      #1{#1}          \fi
\ifx \showarticletitle \undefined \def \showarticletitle #1{#1}   \fi
\ifx \showURL      \undefined \def \showURL       {\relax}        \fi
% invisible to TeX
\providecommand\bibfield[2]{#2}
\providecommand\bibinfo[2]{#2}
\providecommand\natexlab[1]{#1}
\providecommand\showeprint[2][]{arXiv:#2}

\bibitem[cal(2014)]%
        {calcite}
 \bibinfo{year}{2014}\natexlab{}.
\newblock \bibinfo{title}{Apache Calcite}.
\newblock \bibinfo{howpublished}{\url{https://calcite.apache.org/}}.
\newblock


\bibitem[par(2016)]%
        {parquet}
 \bibinfo{year}{2016}\natexlab{}.
\newblock \bibinfo{title}{{Apache Parquet}}.
\newblock \bibinfo{howpublished}{\url{https://parquet.apache.org/}}.
\newblock


\bibitem[arr(2016)]%
        {arrow_ipc}
 \bibinfo{year}{2016}\natexlab{}.
\newblock \bibinfo{title}{Arrow IPC Format}.
\newblock \bibinfo{howpublished}{\url{https://arrow.apache.org/docs/format/Columnar.html}}.
\newblock


\bibitem[grp(2016)]%
        {grpc}
 \bibinfo{year}{2016}\natexlab{}.
\newblock \bibinfo{title}{gRPC}.
\newblock \bibinfo{howpublished}{\url{https://grpc.io/}}.
\newblock


\bibitem[spe(2017)]%
        {spectrum}
 \bibinfo{year}{2017}\natexlab{}.
\newblock \bibinfo{title}{{Amazon Redshift Spectrum}}.
\newblock \bibinfo{howpublished}{\url{https://docs.aws.amazon.com/redshift/latest/dg/c-using-spectrum.html}}.
\newblock


\bibitem[nit(2017)]%
        {nitro}
 \bibinfo{year}{2017}\natexlab{}.
\newblock \bibinfo{title}{{AWS Nitro System}}.
\newblock \bibinfo{howpublished}{\url{https://aws.amazon.com/ec2/nitro/}}.
\newblock


\bibitem[s3s(2017)]%
        {s3select}
 \bibinfo{year}{2017}\natexlab{}.
\newblock \bibinfo{title}{{S3 Select and Glacier Select – Retrieving Subsets of Objects}}.
\newblock \bibinfo{howpublished}{\url{https://aws.amazon.com/blogs/aws/s3-glacier-select/}}.
\newblock


\bibitem[s3(2018)]%
        {s3}
 \bibinfo{year}{2018}\natexlab{}.
\newblock \bibinfo{title}{{Amazon S3}}.
\newblock \bibinfo{howpublished}{\url{https://aws.amazon.com/s3/}}.
\newblock


\bibitem[pre(2018)]%
        {presto}
 \bibinfo{year}{2018}\natexlab{}.
\newblock \bibinfo{title}{{Presto}}.
\newblock \bibinfo{howpublished}{\url{https://prestodb.io/}}.
\newblock


\bibitem[arr(2019)]%
        {arrow_flight}
 \bibinfo{year}{2019}\natexlab{}.
\newblock \bibinfo{title}{Arrow Flight RPC}.
\newblock \bibinfo{howpublished}{\url{https://arrow.apache.org/docs/format/Flight.html}}.
\newblock


\bibitem[aqu(2020)]%
        {aqua}
 \bibinfo{year}{2020}\natexlab{}.
\newblock \bibinfo{title}{{AQUA (Advanced Query Accelerator) for Amazon Redshift}}.
\newblock \bibinfo{howpublished}{\url{https://pages.awscloud.com/AQUA_Preview.html/}}.
\newblock


\bibitem[azu(2020)]%
        {azure-accel}
 \bibinfo{year}{2020}\natexlab{}.
\newblock \bibinfo{title}{Azure Data Lake Storage query acceleration}.
\newblock \bibinfo{howpublished}{\url{https://docs.microsoft.com/en-us/azure/storage/blobs/data-lake-storage-query-acceleration/}}.
\newblock


\bibitem[pol(2022)]%
        {polardbx}
 \bibinfo{year}{2022}\natexlab{}.
\newblock \bibinfo{title}{PolarDB-X}.
\newblock \bibinfo{howpublished}{\url{https://www.alibabacloud.com/product/polardb-x}}.
\newblock


\bibitem[tpc(2022)]%
        {tpch}
 \bibinfo{year}{2022}\natexlab{}.
\newblock \bibinfo{title}{{TPC-H Benchmark}}.
\newblock \bibinfo{howpublished}{\url{http://www.tpc.org/tpch/}}.
\newblock


\bibitem[Abadi et~al\mbox{.}(2008)]%
        {abadi2008column}
\bibfield{author}{\bibinfo{person}{Daniel~J. Abadi}, \bibinfo{person}{Samuel~R. Madden}, {and} \bibinfo{person}{Nabil Hachem}.} \bibinfo{year}{2008}\natexlab{}.
\newblock \showarticletitle{Column-Stores vs. Row-Stores: How Different Are They Really?}. In \bibinfo{booktitle}{\emph{Proceedings of the 2008 ACM SIGMOD International Conference on Management of Data}}. \bibinfo{pages}{967–980}.
\newblock


\bibitem[Charousset et~al\mbox{.}(2016)]%
        {caf}
\bibfield{author}{\bibinfo{person}{Dominik Charousset}, \bibinfo{person}{Raphael Hiesgen}, {and} \bibinfo{person}{Thomas~C. Schmidt}.} \bibinfo{year}{2016}\natexlab{}.
\newblock \showarticletitle{{Revisiting Actor Programming in C++}}.
\newblock \bibinfo{journal}{\emph{Computer Languages, Systems \& Structures}} \bibinfo{volume}{45}, \bibinfo{number}{C} (\bibinfo{year}{2016}), \bibinfo{pages}{105--131}.
\newblock


\bibitem[Chaudhuri et~al\mbox{.}(2004)]%
        {chaudhuri2004estimating}
\bibfield{author}{\bibinfo{person}{Surajit Chaudhuri}, \bibinfo{person}{Vivek Narasayya}, {and} \bibinfo{person}{Ravishankar Ramamurthy}.} \bibinfo{year}{2004}\natexlab{}.
\newblock \showarticletitle{Estimating Progress of Execution for SQL Queries}. In \bibinfo{booktitle}{\emph{Proceedings of the 2004 ACM SIGMOD International Conference on Management of Data}}. \bibinfo{pages}{803–814}.
\newblock


\bibitem[Do et~al\mbox{.}(2013)]%
        {do2013query}
\bibfield{author}{\bibinfo{person}{Jaeyoung Do}, \bibinfo{person}{Yang-Suk Kee}, \bibinfo{person}{Jignesh~M. Patel}, \bibinfo{person}{Chanik Park}, \bibinfo{person}{Kwanghyun Park}, {and} \bibinfo{person}{David~J. DeWitt}.} \bibinfo{year}{2013}\natexlab{}.
\newblock \showarticletitle{Query Processing on Smart SSDs: Opportunities and Challenges}. In \bibinfo{booktitle}{\emph{SIGMOD}}. \bibinfo{pages}{1221–1230}.
\newblock


\bibitem[Duggan et~al\mbox{.}(2011)]%
        {duggan2011performance}
\bibfield{author}{\bibinfo{person}{Jennie Duggan}, \bibinfo{person}{Ugur Cetintemel}, \bibinfo{person}{Olga Papaemmanouil}, {and} \bibinfo{person}{Eli Upfal}.} \bibinfo{year}{2011}\natexlab{}.
\newblock \showarticletitle{Performance Prediction for Concurrent Database Workloads}. In \bibinfo{booktitle}{\emph{Proceedings of the 2011 ACM SIGMOD International Conference on Management of Data}}. \bibinfo{pages}{337–348}.
\newblock


\bibitem[Francisco(2011)]%
        {netezza}
\bibfield{author}{\bibinfo{person}{Phil Francisco}.} \bibinfo{year}{2011}\natexlab{}.
\newblock \bibinfo{title}{{The Netezza Data Appliance Architecture}}.
\newblock
\newblock


\bibitem[Fushimi et~al\mbox{.}(1986)]%
        {grace}
\bibfield{author}{\bibinfo{person}{Shinya Fushimi}, \bibinfo{person}{Masaru Kitsuregawa}, {and} \bibinfo{person}{Hidehiko Tanaka}.} \bibinfo{year}{1986}\natexlab{}.
\newblock \showarticletitle{An Overview of The System Software of A Parallel Relational Database Machine GRACE}. In \bibinfo{booktitle}{\emph{VLDB}}. \bibinfo{pages}{209–219}.
\newblock


\bibitem[Gao and Kozyrakis(2016)]%
        {hrl}
\bibfield{author}{\bibinfo{person}{Mingyu Gao} {and} \bibinfo{person}{Christos Kozyrakis}.} \bibinfo{year}{2016}\natexlab{}.
\newblock \showarticletitle{{HRL: Efficient and Flexible Reconfigurable Logic for Near-Data Processing}}. In \bibinfo{booktitle}{\emph{HPCA}}. \bibinfo{pages}{126--137}.
\newblock


\bibitem[Ghose et~al\mbox{.}(2018)]%
        {ghose2018enabling}
\bibfield{author}{\bibinfo{person}{Saugata Ghose}, \bibinfo{person}{Kevin Hsieh}, \bibinfo{person}{Amirali Boroumand}, \bibinfo{person}{Rachata Ausavarungnirun}, {and} \bibinfo{person}{Onur Mutlu}.} \bibinfo{year}{2018}\natexlab{}.
\newblock \showarticletitle{{Enabling the Adoption of Processing-in-Memory: Challenges, Mechanisms, Future Research Directions}}.
\newblock \bibinfo{journal}{\emph{arXiv preprint arXiv:1802.00320}} (\bibinfo{year}{2018}).
\newblock


\bibitem[Gu et~al\mbox{.}(2016)]%
        {biscuit}
\bibfield{author}{\bibinfo{person}{Boncheol Gu}, \bibinfo{person}{Andre~S. Yoon}, \bibinfo{person}{Duck-Ho Bae}, \bibinfo{person}{Insoon Jo}, \bibinfo{person}{Jinyoung Lee}, \bibinfo{person}{Jonghyun Yoon}, \bibinfo{person}{Jeong-Uk Kang}, \bibinfo{person}{Moonsang Kwon}, \bibinfo{person}{Chanho Yoon}, \bibinfo{person}{Sangyeun Cho}, \bibinfo{person}{Jaeheon Jeong}, {and} \bibinfo{person}{Duckhyun Chang}.} \bibinfo{year}{2016}\natexlab{}.
\newblock \showarticletitle{Biscuit: A Framework for Near-Data Processing of Big Data Workloads}. In \bibinfo{booktitle}{\emph{ISCA}}. \bibinfo{pages}{153–165}.
\newblock


\bibitem[Harmouch and Naumann(2017)]%
        {harmouch2019cardinality}
\bibfield{author}{\bibinfo{person}{Hazar Harmouch} {and} \bibinfo{person}{Felix Naumann}.} \bibinfo{year}{2017}\natexlab{}.
\newblock \showarticletitle{Cardinality Estimation: An Experimental Survey}.
\newblock \bibinfo{journal}{\emph{Proc. VLDB Endow.}} \bibinfo{volume}{11}, \bibinfo{number}{4} (\bibinfo{year}{2017}), \bibinfo{pages}{499–512}.
\newblock


\bibitem[Keeton et~al\mbox{.}(1998)]%
        {keeton1998case}
\bibfield{author}{\bibinfo{person}{Kimberly Keeton}, \bibinfo{person}{David~A Patterson}, {and} \bibinfo{person}{Joseph~M Hellerstein}.} \bibinfo{year}{1998}\natexlab{}.
\newblock \showarticletitle{A Case for Intelligent Disks (IDISKs)}.
\newblock \bibinfo{journal}{\emph{SIGMOD Record}} \bibinfo{volume}{27}, \bibinfo{number}{3} (\bibinfo{year}{1998}), \bibinfo{pages}{42--52}.
\newblock


\bibitem[Kepe et~al\mbox{.}(2019)]%
        {kepe2019database}
\bibfield{author}{\bibinfo{person}{Tiago~R. Kepe}, \bibinfo{person}{Eduardo~C. de Almeida}, {and} \bibinfo{person}{Marco A.~Z. Alves}.} \bibinfo{year}{2019}\natexlab{}.
\newblock \showarticletitle{Database Processing-in-Memory: An Experimental Study}.
\newblock \bibinfo{journal}{\emph{VLDB}} \bibinfo{volume}{13}, \bibinfo{number}{3} (\bibinfo{year}{2019}), \bibinfo{pages}{334--347}.
\newblock


\bibitem[Kim et~al\mbox{.}(2022)]%
        {kim2022learned}
\bibfield{author}{\bibinfo{person}{Kyoungmin Kim}, \bibinfo{person}{Jisung Jung}, \bibinfo{person}{In Seo}, \bibinfo{person}{Wook-Shin Han}, \bibinfo{person}{Kangwoo Choi}, {and} \bibinfo{person}{Jaehyok Chong}.} \bibinfo{year}{2022}\natexlab{}.
\newblock \showarticletitle{Learned Cardinality Estimation: An In-Depth Study}. In \bibinfo{booktitle}{\emph{Proceedings of the 2022 International Conference on Management of Data}}. \bibinfo{pages}{1214–1227}.
\newblock


\bibitem[Koo et~al\mbox{.}(2017)]%
        {summarizer}
\bibfield{author}{\bibinfo{person}{Gunjae Koo}, \bibinfo{person}{Kiran~Kumar Matam}, \bibinfo{person}{Te I}, \bibinfo{person}{H.~V. Krishna~Giri Narra}, \bibinfo{person}{Jing Li}, \bibinfo{person}{Hung-Wei Tseng}, \bibinfo{person}{Steven Swanson}, {and} \bibinfo{person}{Murali Annavaram}.} \bibinfo{year}{2017}\natexlab{}.
\newblock \showarticletitle{Summarizer: Trading Communication with Computing Near Storage}. In \bibinfo{booktitle}{\emph{MICRO}}. \bibinfo{pages}{219–231}.
\newblock


\bibitem[Lee et~al\mbox{.}(2016)]%
        {lee2016operator}
\bibfield{author}{\bibinfo{person}{Kukjin Lee}, \bibinfo{person}{Arnd~Christian K\"{o}nig}, \bibinfo{person}{Vivek Narasayya}, \bibinfo{person}{Bolin Ding}, \bibinfo{person}{Surajit Chaudhuri}, \bibinfo{person}{Brent Ellwein}, \bibinfo{person}{Alexey Eksarevskiy}, \bibinfo{person}{Manbeen Kohli}, \bibinfo{person}{Jacob Wyant}, \bibinfo{person}{Praneeta Prakash}, \bibinfo{person}{Rimma Nehme}, \bibinfo{person}{Jiexing Li}, {and} \bibinfo{person}{Jeff Naughton}.} \bibinfo{year}{2016}\natexlab{}.
\newblock \showarticletitle{Operator and Query Progress Estimation in Microsoft SQL Server Live Query Statistics}. In \bibinfo{booktitle}{\emph{Proceedings of the 2016 International Conference on Management of Data}}. \bibinfo{pages}{1753–1764}.
\newblock


\bibitem[Lin et~al\mbox{.}(2011)]%
        {lin2011llama}
\bibfield{author}{\bibinfo{person}{Yuting Lin}, \bibinfo{person}{Divyakant Agrawal}, \bibinfo{person}{Chun Chen}, \bibinfo{person}{Beng~Chin Ooi}, {and} \bibinfo{person}{Sai Wu}.} \bibinfo{year}{2011}\natexlab{}.
\newblock \showarticletitle{Llama: Leveraging Columnar Storage for Scalable Join Processing in the MapReduce Framework}. In \bibinfo{booktitle}{\emph{Proceedings of the 2011 ACM SIGMOD International Conference on Management of Data}}. \bibinfo{pages}{961–972}.
\newblock


\bibitem[Luo et~al\mbox{.}(2004)]%
        {luo2004toward}
\bibfield{author}{\bibinfo{person}{Gang Luo}, \bibinfo{person}{Jeffrey~F. Naughton}, \bibinfo{person}{Curt~J. Ellmann}, {and} \bibinfo{person}{Michael~W. Watzke}.} \bibinfo{year}{2004}\natexlab{}.
\newblock \showarticletitle{Toward a Progress Indicator for Database Queries}. \bibinfo{pages}{791–802}.
\newblock


\bibitem[Polychroniou et~al\mbox{.}(2014)]%
        {polychroniou2014track}
\bibfield{author}{\bibinfo{person}{Orestis Polychroniou}, \bibinfo{person}{Rajkumar Sen}, {and} \bibinfo{person}{Kenneth~A. Ross}.} \bibinfo{year}{2014}\natexlab{}.
\newblock \showarticletitle{Track Join: Distributed Joins with Minimal Network Traffic}. In \bibinfo{booktitle}{\emph{Proceedings of the 2014 ACM SIGMOD International Conference on Management of Data}}. \bibinfo{pages}{1483–1494}.
\newblock


\bibitem[Raman et~al\mbox{.}(2013)]%
        {raman2013db2}
\bibfield{author}{\bibinfo{person}{Vijayshankar Raman}, \bibinfo{person}{Gopi Attaluri}, \bibinfo{person}{Ronald Barber}, \bibinfo{person}{Naresh Chainani}, \bibinfo{person}{David Kalmuk}, \bibinfo{person}{Vincent KulandaiSamy}, \bibinfo{person}{Jens Leenstra}, \bibinfo{person}{Sam Lightstone}, \bibinfo{person}{Shaorong Liu}, \bibinfo{person}{Guy~M. Lohman}, \bibinfo{person}{Tim Malkemus}, \bibinfo{person}{Rene Mueller}, \bibinfo{person}{Ippokratis Pandis}, \bibinfo{person}{Berni Schiefer}, \bibinfo{person}{David Sharpe}, \bibinfo{person}{Richard Sidle}, \bibinfo{person}{Adam Storm}, {and} \bibinfo{person}{Liping Zhang}.} \bibinfo{year}{2013}\natexlab{}.
\newblock \showarticletitle{DB2 with BLU Acceleration: So Much More than Just a Column Store}.
\newblock \bibinfo{journal}{\emph{Proc. VLDB Endow.}} \bibinfo{volume}{6}, \bibinfo{number}{11} (\bibinfo{year}{2013}), \bibinfo{pages}{1080–1091}.
\newblock


\bibitem[Rescorla(2018)]%
        {tls}
\bibfield{author}{\bibinfo{person}{Eric Rescorla}.} \bibinfo{year}{2018}\natexlab{}.
\newblock \bibinfo{title}{{The Transport Layer Security (TLS) Protocol Version 1.3}}.
\newblock \bibinfo{howpublished}{RFC 8446}.
\newblock
\urldef\tempurl%
\url{https://doi.org/10.17487/RFC8446}
\showDOI{\tempurl}


\bibitem[Riedel et~al\mbox{.}(2001)]%
        {riedel2001active}
\bibfield{author}{\bibinfo{person}{Erik Riedel}, \bibinfo{person}{Christos Faloutsos}, \bibinfo{person}{Garth~A Gibson}, {and} \bibinfo{person}{David Nagle}.} \bibinfo{year}{2001}\natexlab{}.
\newblock \showarticletitle{Active disks for large-scale data processing}.
\newblock \bibinfo{journal}{\emph{Computer}} \bibinfo{volume}{34}, \bibinfo{number}{6} (\bibinfo{year}{2001}), \bibinfo{pages}{68--74}.
\newblock


\bibitem[Stonebraker et~al\mbox{.}(2005)]%
        {stonebraker2005cstore}
\bibfield{author}{\bibinfo{person}{Mike Stonebraker}, \bibinfo{person}{Daniel~J. Abadi}, \bibinfo{person}{Adam Batkin}, \bibinfo{person}{Xuedong Chen}, \bibinfo{person}{Mitch Cherniack}, \bibinfo{person}{Miguel Ferreira}, \bibinfo{person}{Edmond Lau}, \bibinfo{person}{Amerson Lin}, \bibinfo{person}{Sam Madden}, \bibinfo{person}{Elizabeth O'Neil}, \bibinfo{person}{Pat O'Neil}, \bibinfo{person}{Alex Rasin}, \bibinfo{person}{Nga Tran}, {and} \bibinfo{person}{Stan Zdonik}.} \bibinfo{year}{2005}\natexlab{}.
\newblock \showarticletitle{C-Store: A Column-Oriented DBMS}. In \bibinfo{booktitle}{\emph{Proceedings of the 31st International Conference on Very Large Data Bases}}. \bibinfo{pages}{553–564}.
\newblock


\bibitem[Tan et~al\mbox{.}(2019)]%
        {clouddb}
\bibfield{author}{\bibinfo{person}{Junjay Tan}, \bibinfo{person}{Thanaa Ghanem}, \bibinfo{person}{Matthew Perron}, \bibinfo{person}{Xiangyao Yu}, \bibinfo{person}{Michael Stonebraker}, \bibinfo{person}{David DeWitt}, \bibinfo{person}{Marco Serafini}, \bibinfo{person}{Ashraf Aboulnaga}, {and} \bibinfo{person}{Tim Kraska}.} \bibinfo{year}{2019}\natexlab{}.
\newblock \showarticletitle{{Choosing A Cloud DBMS: Architectures and Tradeoffs}}.
\newblock \bibinfo{journal}{\emph{VLDB}} \bibinfo{volume}{12}, \bibinfo{number}{12} (\bibinfo{year}{2019}), \bibinfo{pages}{2170–2182}.
\newblock


\bibitem[Ubell(1985)]%
        {idm}
\bibfield{author}{\bibinfo{person}{Michael Ubell}.} \bibinfo{year}{1985}\natexlab{}.
\newblock \showarticletitle{{The Intelligent Database Machine (IDM)}}.
\newblock In \bibinfo{booktitle}{\emph{Query processing in database systems}}. \bibinfo{pages}{237--247}.
\newblock


\bibitem[Verbitski et~al\mbox{.}(2017)]%
        {aurora1}
\bibfield{author}{\bibinfo{person}{Alexandre Verbitski}, \bibinfo{person}{Anurag Gupta}, \bibinfo{person}{Debanjan Saha}, \bibinfo{person}{Murali Brahmadesam}, \bibinfo{person}{Kamal Gupta}, \bibinfo{person}{Raman Mittal}, \bibinfo{person}{Sailesh Krishnamurthy}, \bibinfo{person}{Sandor Maurice}, \bibinfo{person}{Tengiz Kharatishvili}, {and} \bibinfo{person}{Xiaofeng Bao}.} \bibinfo{year}{2017}\natexlab{}.
\newblock \showarticletitle{{Amazon Aurora: Design Considerations for High Throughput Cloud-Native Relational Databases}}. In \bibinfo{booktitle}{\emph{SIGMOD}}. \bibinfo{pages}{1041–1052}.
\newblock


\bibitem[Verbitski et~al\mbox{.}(2018)]%
        {aurora2}
\bibfield{author}{\bibinfo{person}{Alexandre Verbitski}, \bibinfo{person}{Anurag Gupta}, \bibinfo{person}{Debanjan Saha}, \bibinfo{person}{James Corey}, \bibinfo{person}{Kamal Gupta}, \bibinfo{person}{Murali Brahmadesam}, \bibinfo{person}{Raman Mittal}, \bibinfo{person}{Sailesh Krishnamurthy}, \bibinfo{person}{Sandor Maurice}, \bibinfo{person}{Tengiz Kharatishvilli}, {et~al\mbox{.}}} \bibinfo{year}{2018}\natexlab{}.
\newblock \showarticletitle{Amazon Aurora: On Avoiding Distributed Consensus for I/Os, Commits, and Membership Changes}. In \bibinfo{booktitle}{\emph{SIGMOD}}. \bibinfo{pages}{789--796}.
\newblock


\bibitem[Weiss(2012)]%
        {exadata}
\bibfield{author}{\bibinfo{person}{Ronald Weiss}.} \bibinfo{year}{2012}\natexlab{}.
\newblock \showarticletitle{{A Technical Overview of the Oracle Exadata Database Machine and Exadata Storage Server}}.
\newblock \bibinfo{journal}{\emph{Oracle White Paper.}} (\bibinfo{year}{2012}).
\newblock


\bibitem[Woods et~al\mbox{.}(2014)]%
        {ibex}
\bibfield{author}{\bibinfo{person}{Louis Woods}, \bibinfo{person}{Zsolt Istv{\'a}n}, {and} \bibinfo{person}{Gustavo Alonso}.} \bibinfo{year}{2014}\natexlab{}.
\newblock \showarticletitle{{Ibex: an Intelligent Storage Engine with Support for Advanced SQL Offloading}}.
\newblock \bibinfo{journal}{\emph{VLDB}} \bibinfo{volume}{7}, \bibinfo{number}{11} (\bibinfo{year}{2014}), \bibinfo{pages}{963–974}.
\newblock


\bibitem[Wu et~al\mbox{.}(2013)]%
        {wu2013towards}
\bibfield{author}{\bibinfo{person}{Wentao Wu}, \bibinfo{person}{Yun Chi}, \bibinfo{person}{Hakan Hac\'{\i}g\"{u}m\"{u}\c{s}}, {and} \bibinfo{person}{Jeffrey~F. Naughton}.} \bibinfo{year}{2013}\natexlab{}.
\newblock \showarticletitle{Towards Predicting Query Execution Time for Concurrent and Dynamic Database Workloads}.
\newblock \bibinfo{journal}{\emph{Proc. VLDB Endow.}} \bibinfo{volume}{6}, \bibinfo{number}{10} (\bibinfo{year}{2013}), \bibinfo{pages}{925–936}.
\newblock


\bibitem[Xu et~al\mbox{.}(2020)]%
        {aquoman}
\bibfield{author}{\bibinfo{person}{Shuotao Xu}, \bibinfo{person}{Thomas Bourgeat}, \bibinfo{person}{Tianhao Huang}, \bibinfo{person}{Hojun Kim}, \bibinfo{person}{Sungjin Lee}, {and} \bibinfo{person}{Arvind Arvind}.} \bibinfo{year}{2020}\natexlab{}.
\newblock \showarticletitle{AQUOMAN: An Analytic-Query Offloading Machine}. In \bibinfo{booktitle}{\emph{MICRO}}. \bibinfo{pages}{386--399}.
\newblock


\bibitem[Yang et~al\mbox{.}(2021)]%
        {fpdb}
\bibfield{author}{\bibinfo{person}{Yifei Yang}, \bibinfo{person}{Matt Youill}, \bibinfo{person}{Matthew Woicik}, \bibinfo{person}{Yizhou Liu}, \bibinfo{person}{Xiangyao Yu}, \bibinfo{person}{Marco Serafini}, \bibinfo{person}{Ashraf Aboulnaga}, {and} \bibinfo{person}{Michael Stonebraker}.} \bibinfo{year}{2021}\natexlab{}.
\newblock \showarticletitle{FlexPushdownDB: Hybrid Pushdown and Caching in a Cloud DBMS}.
\newblock \bibinfo{journal}{\emph{VLDB}} \bibinfo{volume}{14}, \bibinfo{number}{11} (\bibinfo{year}{2021}), \bibinfo{pages}{2101–2113}.
\newblock


\bibitem[Yu et~al\mbox{.}(2020)]%
        {pushdowndb}
\bibfield{author}{\bibinfo{person}{Xiangyao Yu}, \bibinfo{person}{Matt Youill}, \bibinfo{person}{Matthew Woicik}, \bibinfo{person}{Abdurrahman Ghanem}, \bibinfo{person}{Marco Serafini}, \bibinfo{person}{Ashraf Aboulnaga}, {and} \bibinfo{person}{Michael Stonebraker}.} \bibinfo{year}{2020}\natexlab{}.
\newblock \showarticletitle{PushdownDB: Accelerating a DBMS using S3 Computation}. In \bibinfo{booktitle}{\emph{ICDE}}. \bibinfo{pages}{1802--1805}.
\newblock


\end{thebibliography}

\end{document}